# Practopoiesis: Or how life fosters a mind

Danko Nikolić

- Department of Neurophysiology, Max Planck Institute for Brain Research, Deutschordenstraße 46, D-60528 Frankfurt/M, Germany

- Frankfurt Institute for Advanced Studies (FIAS), Ruth-Moufang-Straße 1, D-60438 Frankfurt/M, Germany

- Ernst Strüngmann Institute (ESI) for Neuroscience in Cooperation with Max Planck Society, Deutschordenstraße 46, D-60528 Frankfurt/M, Germany

- Department of Psychology, Faculty of Humanities and Social Sciences, University of Zagreb, Croatia

Correspondence:
Danko Nikolić
Max-Planck Institute for Brain Research
Deutschordenstr. 46
60528 Frankfurt am Main
email: danko.nikolic@gmail.com
www.danko-nikolic.com

**Abstract**

**The mind is a biological phenomenon. Thus, biological principles of organization should also be the principles underlying mental operations. Practopoiesis states that the key for achieving intelligence through adaptation is an arrangement in which mechanisms laying a lower level of organization, by their operations and interaction with the environment, enable creation of mechanisms lying at a higher level of organization. When such an organizational advance of a system occurs, it is called a traverse. A case of traverse is when plasticity mechanisms (at a lower level of organization), by their operations, create a neural network anatomy (at a higher level of organization). Another case is the actual production of behavior by that network, whereby the mechanisms of neuronal activity operate to create motor actions. Practopoietic theory explains why the adaptability of a system increases with each increase in the number of traverses. With a larger number of traverses, a system can be relatively small and yet, produce a higher degree of adaptive/intelligent behavior than a system with a lower number of traverses. The present analyses indicate that the two well-known traverses—neural plasticity and neural activity—are not sufficient to explain human mental capabilities. At least one additional traverse is needed, which is named *anapoiesis* for its contribution in reconstructing knowledge e.g., from long-term memory into working memory. The conclusions bear implications for brain theory, the mind-body explanatory gap, and developments of artificial intelligence technologies.**

## 1. Introduction

To help solve the brain-body problem (Descartes 1983/1644; Popper 1999; Chalmers 1999; Rust 2009), systems neuroscience needs to near-decompose (Simon 1994; Bechtel and Richardson 1992)[1] the complex biology of the brain into simple components. Likewise, biology is still in a need of a general theory of interactions that would explain relationships between its different levels of organization (Noble 2008a, 2008b; Bateson 2004). The present work is an attempt to develop a theory that satisfies both of these needs.

The heart of the present approach can be illustrated through the role that plasticity mechanisms play in neural networks. Be it a biological network or one simulated on a computer, without plasticity mechanisms, it would be impossible to endow the network with the structure necessary to accomplish its tasks across different environments if each environment poses different demands. Plasticity mechanisms are the means of steering the network into the desirable state of operation. Once created, the network offers another mechanism of equal steering importance: neural activity. The muscles and skeleton of a body provide machinery to generate movement and behavior. But they are useless without a network of neurons, which controls those movements. Neurons with their electro-chemical activity, and through inhibition/excitation, steer effectors and ultimately give life to the motion of the body. The present approach emphasizes that what plasticity is for a network, the network is for behavior: In both cases there is an enabling force. Both forces need to work well, and they lie in an organizational hierarchy: The rules of plasticity are organizationally lower than network anatomy, and anatomy is organizationally lower than the generated behavior. It is always that higher levels are a result of operations of lower levels and not the other way around.

The present work generalizes this lower-to-higher relationship and proposes a formal theory. This makes it possible to ask what happens if there are not only two adaptive mechanisms (plasticity and neural activity) but three (e.g., in case of an organism and its brain and mind) or four (in case of evolution operating on an entire species). Would more levels produce more intelligent behavior and how many levels are really used by biological systems?

The theory is named *practopoiesis*—derived from Ancient Greek words πρᾶξις (praksis), meaning "action, activity, practice" *and* ποίησις (poiesis), from ποιέω (poieo), "to make". The term *practopoiesis* refers to "creation of actions", emphasizing the fact that physiological mechanisms at any level of adaptive organization operate through actions—and requires mechanisms to be put in place capable of executing those actions. For example, gene expression mechanisms act, plasticity mechanisms act, and neurons act. All of those mechanisms need to be properly created. The name practopoiesis is also a tribute to the theory of autopoiesis (Maturana & Varela 1980, 1992), which is one of the precursors to the present work—providing the insights that the process of creating new structures, or *poiesis*, in biological systems underlies both the physiology of an organism and its mental operations (see also Thompson 2007).

Another theory that is a precursor to practopoiesis is self-organization (Ashby 1947; Andrew 1979; Szentagothai and Erdi, 1989). Practopoiesis can be considered a specific implementation of the principles of self-organization namely those applicable to biological processes. While self-organization generally applies also to non-biological phenomena (such as galaxies or chemical reactions), practopoiesis applies exclusively to adaptive systems. This allows practopoiesis to specify more accurate the organization principles, which happen to be founded in several theorems fundamental to cybernetics (Ashby 1958; Conant & Ashby 1970). Being less general than self-organization practopoiesis applies only to a subset of all the processes of self-organization that take place within an organism. Practopoiesis is concern only with those aspects that enable the organism to adapt to newly emerging circumstances in its environment. For example, retino-topic mapping between the eye and the structures in the central nervous system is established through self-organization. However, some of these processes may occur with intrinsic rhythms only, without a need for inputs from the environment (Szentagothai and Erdi, 1989; Eglen and Gjorgjieva, 2009). Others require adaptive changes made in response to the current of the environment (e.g., Singer and Tretter 1976; Singer et al., 1981). Practopoiesis applies only to the latter ones. The explanatory power that practopoiesis brings to the problem of mind and brain is the suggestion that proposed at the end of this manuscript stating that mental operations are by their nature adaptive. Hence, the principles of practopoiesis provide the right tools for understanding how, through self-organization of the system, the mental emerges out of physical.

## 2. Practopoiesis: A general theory of adaptive systems

One of the key postulates of practopoiesis is the necessity of interactions with the environment. The idea is that each adaptive mechanism, at any level of organization, receives its own feedback from the environment. That way, practopoiesis follows the traditions of and is in a general agreement with the ecological approach to mental operations (Gibson 1977, 1979), enactivism (Varela et al. 1991; Noë 2012), externalism (Holt 1914; Brooks 1991) and other works concerning situated and embodied cognition (e.g., Lakoff & Johnson 1980; Damasio 1999; McGann et al. 2013; Di Paolo & De Jaegher 2012), and robotics (Brooks 1999). Also, various preceding works considering feedback interactions (Friston 2010; Shipp et al. 2013; Friston et al. 2012; Bernstein 1967; Powers 1973) provide important background for the present work.

Practopoiesis can be fundamentally considered a cybernetic theory. Cybernetics studies control systems based on feedback loops (e.g. Wiener 1961) (Figure 1A). Practopoiesis is grounded in the theorems of cybernetics—foremost, the law of requisite variety (Ashby 1958; Beer 1974, 1979) and the good regulator theorem (Conant & Ashby 1970). Practopoiesis is an extension in a sense that it explains how systems obtain their cybernetic capabilities i.e., how they learn what and where to control. Hence, practopoiesis can be understood as a form of a second-order cybernetics, or cybernetics-of-cybernetics (Heylighen & Joslyn 2001; Glanville 2002; von Foerster 2003).

### *2.1 Three main telltale signs of practopoietic systems*

To determine whether a system has the capability to learn to control these properties must be observed:

1) *Monitor-and-act machinery*: An adaptive system must consist of components that are capable of detecting conditions for a necessity to act, and of acting. These components monitor their own surrounding world, make changes through actions, and then evaluate the effects e.g., determine whether further actions are needed. For example, gene expression mechanisms constitute one type of monitor-and-act units. Neurons constitute another type of monitor-and-act units.

2) *Poietic hierarchy*: The monitor-and-act units are organized into a hierarchy in which low-level components, by their actions, create, adjust, service and nourish high-level components. Once created, higher-level components can in principle operate on their own. That is, at least for some time, they do not require further engagements of the lower-level components. This is achieved by creating new physical structures that constitute higher-level monitor-and-act units. For example, gene expression mechanisms create neurons and determine their properties, but not the other way around.

3) *Level-specific environmental feedback*: Monitor-and-act components receive necessarily feedback from the environment to which the system is adapting. This means that units at different levels of organization receive different type of feedback. For example, neurons constituting a patellar reflex receive different type of feedback from the environment than the gene expression mechanisms that build that reflex mechanism on the first place.

These properties can be illustrated by a simple interaction graph (Figure 1B): The monitor-and-act units operating at the top of the hierarchy can be described as classical cybernetic systems (as in Figure 1A). However, other units, lower on the hierarchy, add complexity to the system. These units monitor the effects that the top of the hierarchy produces on the environment and, when necessary, make alterations. For as long as higher-level components satisfy the needs of an organism, there will be no need for changes at lower levels of system organization. But if higher-level components are unsuitable, they are being poietically adjusted. For full functioning, two types of feedback from the environment are required, one for each level (Figure 1B). In case that a low level fails to receive feedback from the environment but instead receives feedback only from within the system, the system's capability to adapt to the environment at that level of organization is lost. In that case no separate levels of practopoietic organization can be claimed.

### *2.2 The main desideratum: Cybernetic knowledge*

To work properly and harmoniously with an environment, every component of a system must be adjusted according to its environment. The proper adjustment can be referred to as *cybernetic knowledge* of that component e.g., knowledge on when to act and how (Ashby 1958). Cybernetic knowledge is necessarily subjected to Conant & Ashby's good regulator theorem (Conant & Ashby 1970), stating: "any successful control mechanism must be a model of the system that it controls". That is, one can deal with the surrounding world successfully only if one already possesses certain knowledge about the effects that one's actions are likely to exert on that world[2]. Maturana and Varela (1980, 1992) expressed it as: "All doing is knowing and all knowing is doing."

The combination of poiesis and level-specific environmental feedback has the following implication: The process of building the system is also the process of adapting the system, which is also the very process of acquiring cybernetic knowledge. Building a system through interaction with an environment and adjusting to it cannot be distinguished from acquiring cybernetic knowledge about this environment. That way, newly created structures become a model (Conant & Ashby 1970) of the system's environment. For example, variation in phenotype for the same genotype (Johanssen 1911) is a form of practopoietic extraction of knowledge.[3] Formation of neural network architecture (e.g., synaptic connectivity) through interactions with the environment is also a form of extraction of cybernetic knowledge.

### *2.3 Knowledge requires variety*

The total amount of cybernetic knowledge deposited within a system is related to the total number of different states that the system can assume while interacting with the environment, and is referred to as the *cybernetic variety* of the system. The demands on variety are determined by Ashby's *law of requisite* variety (Ashby 1958; Beer 1974, 1979), which states that for a successful control of a system, the system that controls has to have at least as many states as the system being controlled. Thus, being a good model of the environment entails a sufficient number of states, which is a pre-requirement to store a sufficient amount of cybernetic knowledge within the systems.

### *2.4 Practopoietic transcendence of knowledge: Generality-specificity hierarchy*

The contribution that practopoietic theory brings on top of the existing cybernetic theory is the introduction of the adaptive hierarchy. This hierarchy implies a specific relation between the cybernetic knowledge that drives a poietic process and the knowledge that has been extracted through that process: The knowledge that can be instilled at a new level of organization is always limited by what the system had known prior to the process of poiesis: higher-level knowledge is always a specific case of lower-level knowledge. Kant referred to this limitation as transcendence of knowledge (Kant 1998). In machine learning, this system property is known as inductive bias (Mitchell 1980).

A higher level of organization contains knowledge about how the environment has previously responded to the actions at lower levels. Consequently, the relationship between knowledge levels can be always described as a change in knowledge specificity along the organizational hierarchy. Knowledge at a higher-level system organization must always be a specific case of more general knowledge at a lower level of organization.[4]

This relation can be shown even in the simplest, non-biological forms of cybernetic systems. For example, a thermostat with a sensor and a heater can be deemed a simple monitor-and-act unit possessing cybernetic knowledge on how to keep a space comfortably warm. This unit has two levels of organization, general and specific: The general knowledge of that system can be expressed as a relation between the *input* (current temperature) and the *output* (heating intensity). For example, *output* = (*target* – *input*) / 3. Specific knowledge is then derived by the actions of this controller. For example, specifically, right now *input* may be 35, and *target* may be 20. The needed *output* is thus -5.

In biology, an example of the generality-specificity relation is the general rule about when and which proteins should be synthesized versus the specific proteins that have been synthesized and resulted in a certain phenotype. The latter reflects the properties of a particular environment within which the system operated recently, whilst the former reflects the properties of the environment across a range of time and space covered by the evolution of the organism. Thus, a phenotype will always contain more specific knowledge than the genotype.

The generality-specificity relationship applies not only to the gene-to-protein relationship but also to higher levels of system organization. The anatomical connectivity of a nervous system reflects more general cybernetic knowledge than that of neuronal activity: The anatomy contains knowledge on what to do in general, across a range of sensory inputs, whilst the current electrical state of the network contains the knowledge of what is going on right now.

The graphs of interactions within real cybernetic systems (e.g., Figures 1A, B) can be quite complex if the variety of the system is large. Hence, *knowledge graphs* can be introduced, with which the essential practopoietic relationships between different levels of knowledge organization can be illustrated. Figure 1C-left illustrates the simplest knowledge graph for a control system (e.g. a thermostat): architecture is more general than the current state. The knowledge provided in the form of architecture is used to extract more specific knowledge in a form of system state (the sizes of the circles in knowledge graphs can be used to indicate the relative variety at each level).

But this needs not be the limit. Additional mechanisms may increase the adaptability in a form of adjusting system architecture (Figures 1B and 1C-right). Such low-level adjustments of architecture using environmental feedback are known as supervised learning e.g., the back-propagation algorithm (Rummelhart et al. 1986) (illustrated in Figure 2A as an interaction graph), or as activity-dependent plasticity (Dubner & Ruda 1992; Ganguly & Poo 2013) (illustrated in Figure 2B as a knowledge graph). In either case, the system has three levels of organization.

In Figure 2, the knowledge stored in the rules of the plasticity mechanisms lies at the lowest level of organization. The application of these rules leads to extraction of new knowledge at the anatomical level. The application of anatomical knowledge leads to the extraction of new knowledge at the highest level of organization—the activity of neurons and consequent generation of input-output interactions i.e., behavior. Thus, ultimately, every behavioral act is a specific expression of the general knowledge stored in our learning mechanisms (i.e., our genes). Our genes know what is good for our survival and proliferation in general. Our behavioral acts know what should be done specifically in a given situation—right now.

In conclusion, the set of all kinds of specific knowledge that a system can possibly learn is limited by the general knowledge that a system begins its life with. One cannot learn specifics for which one has not already pre-evolved a more general learning system. Ultimately, every skill that we acquire and every declarative fact we memorize is a specific form of general knowledge provided by our genes (e.g. Baum 2004).

*2.5 Traverse is a generator of variety*

The introduction of the practopoietic hierarchy implies that the transition from high to low generality of knowledge is an active process. We refer to this process here as an *adaptive traverse of knowledge*, or simply a *traverse*. A traverse is a process, or a set of operations, by which changes are made through system's interaction with the environment such that the system has acquired new operational capabilities, or has directly adjusted its environment to its needs. Whenever a system operates i.e. its monitor-and-act units are engaged, the system executes a traverse. The number of traverses that a system possesses and executes depends on its organization. This number is affected primarily by the number of levels at which the system interacts with its environment, which in turn determines the number of levels at which that system makes changes to itself. Systems with one traverse interact with the environment directly and make no changes to themselves. An additional traverse is needed to make changes to itself in which case the interaction with the environment becomes indirect—through the functioning of the other traverses that the self-changes have affected. For the system to be adaptive each of the traverses has to receive feedback from the environment—and each feedback has to provide a specific type of information, relevant for the monitor-and-act units of this particular traverse. Whenever there is more than one traverse, the lower one on the hierarchy always determines how the upper one operates. Formally, we can define a traverse as a process in which more general cybernetic knowledge has been used throughout the operations of the system to extract more specific cybernetic knowledge. Note that the total number of levels of organization that possess cybernetic knowledge is always larger by one than the number of traverses within the system (indicated by arrows in knowledge graphs). This is because the top level of organization produces output[5], which affects the environment instead of poietically affecting the system[6].

An example of a traverse is when the general knowledge of network plasticity mechanisms—about when and what to change anatomically—, creates new functional capabilities—e.g., on when and how to respond to sensory stimuli. Yet another traverse at a higher level of organization occurs when this network operates by closing sensory-motor loops and creating behavior. In both examples, more general knowledge is applied to create more specific one. One more example of a traverse is when gene expression mechanisms, under the influence of environmental factors, generate anatomical structures. Here, gene expression fosters new functional capabilities for the organism. A biological system undergoes a traverse also through operations of its organs. A digestive tract has an important enabling role for the organism. And so does the immune system, which, with its operations, realizes new, functionality: a healthier state of the organism. Also, Darwinian evolution by natural selection can be considered a traverse when an entire species is considered a system. Evolution is the lowest-level traverse known in biological systems. [7]

In general, a traverse is when more general cybernetic knowledge of monitor-and-act units is used to produce certain beneficial effects for the system in a form of implementing new, more specific cybernetic knowledge. The latter is then considered higher on the organizational hierarchy than the former.

Thus, creating new structures is equivalent to the system's adaptation, which is equivalent to extracting cybernetic knowledge, which can be expressed as a traverse from general to specific knowledge. A traverse is the central adaptive act of a practopoietic system.

Traverse is also how a system generates cybernetic variety. A small number of general rules can be used to extract a large number of specific ones. An example are artificial neural networks.[8]

The total number of traverses matters. Some systems have a single traverse (e.g., thermostat, cybernetic feedback loop in Figure 1A), while others have more than one traverse (e.g., living systems, neural networks; Figure 2). Importantly, addition of one more traverse provides the system with much more capability to generate variety—even when the system is leaner: One system may use huge resources to store all actions for all situations that could possibly be encountered. Another system that has an additional traverse may compress that knowledge to a few general rules and infer in each situation the relevant actions.[9],[10] The latter one is more adaptive and yet leaner. We can say that each additional traverse provides a *variety relief* for the system.

Traverses are crucial for near-decomposability of complex adaptive systems into components relevant for understanding how the system works.

### *2.6 Level-specific environmental feedback and practopoietic cycle of causation*

In practopoietic systems, interactions need to close the causal chain of events through the highest level of (self-)organization. Evolution does not know whether a change is good until a full-fledged organism is developed to interact with the environment. This requires involvement at the top of the hierarchy i.e., behavior. Similarly, genes do not fully know which proteins to synthesize until the organism interacts with the environment at the highest level of organization and probes the environment.

Thus, the feedback loop is closed by generating behavior and then getting feedback on the effects that this behavior exerted. This follows from the poietic properties of systems: Actions of low-level mechanism produce effects on higher-level mechanisms, which then produce effects on the environment (Figure 1B). In fact, in practopoietic systems, there is no way around this involvement of the top. If the causality flowed in any other way, a shortcut would have been found to affect the environment directly, without the higher levels of organization. The system would maybe act faster, but would lose its adaptive capabilities, the degree of loss corresponding to the number of organization levels skipped due to the shortcut.

Thus, as illustrated in Figure 1B, upward causation should occur within the system, and this is a process of poiesis. In contrast, downward causation should take the path outside the system and through *level-specific environmental feedback*. This is the only way for the poietic process to receive feedback from the environment, and for the system as a whole to extract cybernetic knowledge and become a good regulator. For example, a lack of certain nutrients may cause the expression of certain genes, which may be in turn responsible for plastic changes in the nervous system. These changes can then affect behavior patterns in such a way that the organism successfully obtains the needed nutrients, which eventually ceases the expression of the said genes. This entire loop of internal upward poiesis and external downward feedback through multiple levels of organization is referred to as the *practopoietic cycle (or loop) of causation*.

### *2.7 Equi-level interactions*

In any given adaptive system the total number of practopoietic levels of organization is likely to be smaller than the total number of monitor-and-act units of that system. For example, so far, we discussed three possible traverses of knowledge relevant for a nervous system—based respectively on neural activity, plasticity, and evolution. In contrast, a nervous system consists of many billions of monitor-and-act units that take many different physiological forms. Therefore, many units will not be related hierarchically, but will operate at the same level of organization and thus, will undergo *equi-level interactions*.[11]

Equi-level interactions occur when two or more components do not exhibit all the properties required for adaptive practopoietic organization: monitor-and-act units, poietic hierarchy and level-specific feedback. There are several scenarios in which two interacting units, $A$ and $B$, violate the requirement for practopoietic (self-)organization:

2.7.1) *Environmental interactions*: Interactions in neither of the two directions ($A$ to $B$ or $B$ to $A$) occur directly within the system but instead all interactions occur through the environment. This occurs when two units lack internal means of interaction and yet affect the overall behavior of the system. For example, genes in one skin cell and genes in another neural cell do not have means to interact directly. Nevertheless, they interact because each affects somehow the input that the other receives from the environment: The actions of a neuron may affect behavior, which in turn affects the amount of sun exposure that the skin cell receives, which in turn affects its gene expression. Conversely, the color of skin determined by that cell may affect how the social environment responds to an individual, which then affects the inputs to that neuron. Similarly, two neurons may interact chiefly through the environment; A motor neuron in the spinal cord may induce body movements that change the image projections on the retinae, affecting hence the activity of the neurons in the visual system. For those cells, monitor-and-act units from all the levels of organization (i.e., *top*-1, *top*-2, etc. in Figure 1C; read as "top minus one", "top minus two", etc.) may interact through the environment. This includes plasticity mechanisms of different cells (e.g., Yoshitake et al. 2013)[12].

These interactions through the environment are probably the most common form of equi-level interactions in biological systems. Therefore, we can say that adaptive systems are largely *environmentally interactive*—i.e., their components interact by closing the practopoietic cycle of causation.

The top level of organization with its output functions into the environment is the glue that puts the interactions among all of the components of the system together. By relying on such indirect interactions, the system's knowledge can grow linearly with its physical size; new monitor-and-act units can be added without having the burden of implementing the hardware for interaction pathways, the combinatorics of which grows faster than linearly. The organism's interaction with the environment does the "connecting" job.

2.7.2) *Bi-directional interactions*: Direct physical interactions occur within the system, but in both directions (from $A$ to $B$, and from $B$ to $A$). Hence, in this case no unit can be identified as higher or lower on the poietic hierarchy—all units possess equal level of generality (specificity) of cybernetic knowledge. Examples of such interactions are neurons connected into a nervous system, which exhibit largely reciprocal, reentrant connections (direct or indirect). Thus, as soon as one neuron attempts to assume higher generality of cybernetic knowledge by controlling others, the following reentrant input controls this neuron in return, and by doing so restores the generality status of its cybernetic knowledge back to the same level as the others. (For the requirements to successfully separate the levels of generality see Knowledge shielding bellow.)

2.7.3) *Feedback-less interactions*: Direct physical interactions among components only occur in one direction but there is no feedback from the environment that would uniquely provide adaptive capabilities to one of the components and thus, distinguish it form from others. In this scenario, the chain of causal events within the system is fully determined by the internal physical properties of that system and the outcome is not controlled by the properties of the environment. For example, the process that begins with DNA transcription and ends with a synthesis of a protein has multiple stages but the environment may not regulate these stages—once the transcription has started it reliably ends always with the same protein. Similarly, execution of a simple reflex has many subcomponents (depolarization, action potential generation, neurotransmitter release, muscle contraction) but once the events started unfolding, they may not be any longer adjusted by the feedback from the environment. In both cases, the monitor-and-act units constituting the chain of events operate all at the same level of organization.

### *2.8 Downward pressure for adjustment*

Understanding conditions that initiate changes to the system at the low levels of organization is important for understanding adaptive practopoietic systems, which is a problem related to the issue of downward causation (Noble 2008a, 2008b; Bateson 2004; Campbell 1990; Bedau 2002). In practopoiesis, effects towards down occur through level-specific environmental feedback. The top level of organization acts on the environment, and then the environment informs lower levels that the higher ones may not have performed their jobs successfully. That is, the signal for a need to act at lower levels is an event that has both of the following properties: i) it has been established in the past that this signal indicates a need for action, and ii) higher levels did not manage, for whatever reason, to eliminate that need (i.e., eliminate the signal).

In that case, through level-specific environmental feedback, the system experiences a *downward pressure for adjustment*: Changes are needed at lower levels of system organization in order to change—adaptively—the properties of the higher levels. In other words, by actions of monitor-and-act units laying at the bottom of the hierarchy a new system with new cybernetic knowledge is created at the top. For example, various metabolic indicators during a cold season affect gene expression such that an animal grows thicker fur; or changes in gene expression due to chronic malnutrition create behavioral changes that force an animal to change its habitat. A species may find itself under a (downward) pressure to evolve.

As a result of such adaptive capabilities, the total variety of the system's interactions with the environment is much higher when observed across different demands from the environment, than when the system is observed within relatively stable environmental conditions.

Downward pressure for adjustment always involves the environment and is often induced by novelties in the environment. In stable environments, low-level mechanisms experience little pressure for change. Downward pressure for adjustment triggers a practopoietic cycle of causation and thus, involves actions at the higher levels of organization. The changes made to higher levels often cannot be made quickly because there is no direct instruction on how to fix the problem. Low-level cybernetic knowledge gives certain strategies on how to approach the problem, but often does not give a direct solution. In those cases the solution is approached iteratively: The system must make one change and test the success of that attempt, and if it was not sufficient it may need to make another, maybe a different attempt, testing it again and so on.[13]

Downward pressure is exerted on neuronal plasticity mechanisms to adjust the anatomy of the system as a result of changes in the environment (new events). Every form of learning is a result of the pressure to fix discrepancies between the existing sensory-motor operations and those required by the surrounding world. The downward pressure is on making more efficient behavioral actions, percepts, memory recalls, etc. Similarly, evolution by natural selection can be under more or less pressure for change, when organisms are more or less adapted to the given environment. In either case, it is chiefly the environment that dictates when changes need to be made.

*2.9 Trickle-down information*

Another issue related to downward causation is the question of information traveling in downward direction. A general question is: How do monitor-and-act units at lower levels of organization obtain information from environment if information can enter only through the top level? And if they receive such information, is it based on a form of downward causation? Examples are various: Learning mechanisms in neural networks use extract feedback from the very same sensory inputs that drive sensory-motor loops; Gene expression may be regulated by concentration of molecules, which in turn may depend on feeding behavior of the animal.

Practopoiesis presumes a general property of adaptive systems: While such systems interact with the environment, they necessarily acquire information about their environment. Consequently, their internal structure and dynamics already contains information available to monitor-and-act units operating at various levels of organization. This property is called *trickle-down information* and follows from the Conant and Ashby's (1970) good regulator theorem: If the various components of the system have been adapting the system to the environment in recent past, the system itself must be informative of that environment. It is then a question of equipping monitor-and-act units with a proper cybernetic knowledge to use that information.

Another example of trickle-down information is the mechanical stress produced by the musculoskeletal system, which in turn evokes physiological responses in the nervous system—process known as neurodynamics (Shacklock 1995). In neurodynamics, high-level mechanisms induce behavior that provides, through mechanical causal pathways, inputs to low-level physiological mechanisms. Adaptive functions of these mechanisms have often valuable effects for the organism as a whole, as observed e.g. through the benefits of physical exercise for one's health.

Information that travels practopoietically downwards must be based on physical interactions and thus, on causal events. Importantly, however, this causation is not poietic. The information that arrives down only produces pressure for adjustment. In other words, these downward-headed physical events only serve as triggers for activating (or deactivating) poietic processes executed by monitor-and-act units. These poietic actions then only exert effects towards up. These non-poietic trickle-down causal effects may be considered as "side-effects" exerted towards down while high-levels units execute their poietic functions toward up. Low-level units learn to take advantage of these side effects. One reason that poietic effects cannot occur towards down is the need to shield cybernetic knowledge.

*2.10 Knowledge shielding*

From transcendence of knowledge (i.e., from inductive bias) follows a need to ensure resilience of knowledge at lower levels of organization to events that occur at higher levels of organization. The short-lived specific pieces of cybernetic knowledge supersede each other with high pace. They should not spoil the general knowledge acquired through much longer periods of time and reflecting much more "wisdom". For example, the current extreme temperature to which a thermostat may be exposed should not affect the general input-output functions of the thermostat; A temporary event of a low air temperature should not immediately cause the animal to change its genome into growing a thick fur. A search for a shelter may be a much more appropriate function.

In other words, the general knowledge at lower levels of organization should be protected, or *shielded*, from the whims of the specific events at higher levels of organization. The uni-directional flow of poiesis in adaptive systems serves exactly this purpose: This organization ensures that general knowledge makes decisions about specific knowledge, but not the other way around; The "big picture" must not be muddled by the current affairs. This separation of knowledge is the key to success of an adaptive system.

If knowledge shielding would be suddenly lost and the current events at the high levels of organization would readily change the knowledge at lower levels of organization—e.g., if the behavior of an animal could change animal's genes—, the adaptive capabilities of the system would quickly fall apart. The system would not be able to adapt any longer—or even survive. The general knowledge in our genes, acquired through eons, would quickly water down by the current events. The organism would be adapted only to very specific most-recent circumstances, and would lose the knowledge needed to the deal with a variety of circumstances, which the ancestors have painstakingly acquired through a long process of natural selection. Cybernetic wisdom would be replaced by folly.

Similarly, if knowledge shielding was suddenly lost within a nervous system, memories and skills acquired throughout lifetime could vanish in a wake of a single novel event: One may forget how to walk while one swims.

The historical debate between Charles Darwin's natural selection and Jean Lamarck's acquired characteristics was in fact a debate of whether inherited information is shielded. Darwin's theory presumed knowledge shielding whereas Lamarck's presumed the opposite—a system without a practopoietic hierarchy, in which a single giraffe could extend its neck by interacting with unusually high trees and then transfer this new knowledge to its ancestors. By now, much evidence has been accumulated to indicate that largely Darwin was right and that Lamarckian type of inheritance plays a very small role, if any. Practopoiesis explains why this has to be the case: Lamarckian approach would be disastrous for the knowledge acquired by distant ancestors. If just one or few generations experienced absence of a certain food source or a certain predator, the knowledge on how to find, consume and digest that food, or how to recognize, avoid and fight that predator may wash out, having disastrous effects on the species as a whole.

Physiological mechanisms by which living cells shield genetic knowledge are epitomized in Crick's (1958; 1970) *central dogma of molecular biology*: Information can be transferred from DNA to proteins but not the other way around. Currently, much less is known about the mechanisms that shield memories within the nervous system.

### *2.11 Intelligence: Traverses combined with variety*

Another factor that enhances the adaptability of a system is addition of a traverse. To analyze the possible limitations of the current brain theory and artificial intelligence (AI) algorithms, and to determine whether they can be improved by addition of a traverse, it is necessary to establish what additional traverses bring to the overall system's intelligence. We have made a case that adding a level of organization gives more adaptive advantages to a system. For example, adding organization levels at the bottom of the hierarchy, in form of evolution, can be useful. But adding organizational levels in the middle of the hierarchy can help adaptability too. A system that evolves an intermediate adaptive stage is more adaptive than a system lacking that stage. For example, genes do not act directly on the environment but create nervous system, which then acts. Here, the nervous system plays a role of an intermediate adaptive stage. In addition, a network equipped with plasticity rules is more adaptive then the network without plasticity rules. Also, plasticity rules may be adjusted and produce even a more adaptive system[14].

Intermediate adaptive stages provide the space needed to adjust system's own properties i.e., to learn. A system that is powerful in acting on the environment but is unable to act on "itself" is much less adaptive than a system that is able to change its own structure.

Practopoietic theory emphasizes the importance of each additional traverse of a system. The more organizational levels spanned by traverses, the better the coverage of the generality-specificity continuum of cybernetic knowledge. Thus, a system may posses a large amount of knowledge, but yet may not be very adaptable, much like a book may contain much information and still be unable to exhibit intelligence and rewrite itself because it has no traverses. In contrast, a thermostat has one traverse and although it deals with only one variable at a time (it has low variety), in terms of practopoiesis, it is more adaptive than a book. A thermostat, with its traverse, has one more form of interaction with the environment than a book. Similarly, a computer may store and process more information than the genome of the simplest bacteria (gigabytes as opposed to 1.3 megabytes of Pelagibacter ubique; Giovannoni et al. 2005) and yet, due to its multiple traverses, a bacterium is a system of a higher degree of adaptability than a computer.

Despite this increase in adaptive levels in biological systems, their total adaptive power i.e., their intelligence, is given by a combination of the number of traverses and the total cybernetic variety possessed by the system. The systems that posses the same number of traverses are set apart by the amount of cybernetic knowledge. Additional knowledge can increase richness of behavior too. For example, a Braitenberg vehicle (Braintenberg 1984) consisting of two controllers and can produce much richer dynamics than a single controller of a thermostat, and hence may exhibit higher intelligence. And the larger human genome can produce more than the small genome of bacterium Pelagibacter ubique (~750 megabytes vs. 1.3 megabyte). Similarly, a human brain can produce richer behavior than a mouse brain due to the variety produced by the total number of cells.[15]

Nevertheless, the number of traverses makes a crucial difference for how a system can generate variety. With loss of each traverse more variety must be pre-stored, and the future needs already need to be known at the time of storage. In contrast, with an additional traverse, the variety can be generated and adjusted as events unfold. The process of extraction of cybernetic knowledge ensures that the organism continues to be a good model of its environment. Hence, systems with a larger number of traverses can be smaller in total size and yet, produce the same or higher amount of variety than systems with a smaller number of traverses. This has critical consequences for operations in environments unpredictable at a certain shorter time-scale—those who's recent past is not necessarily a good predictor of immediate future. The less predictable the surrounding world at short time scales is i.e., the more the long-term statistics needs to be considered, the higher the advantage of an additional traverse. An additional traverse may enable the system how to adjust more efficiently to new circumstances in the surrounding world.

This brings us to a realization that all cybernetic knowledge must have a source, i.e. a level below that has extracted it. Knowledge of biological systems can be tracked down to Darwin's evolution by natural selection i.e., to the most fundamental piece of knowledge of all: It is good for the species to make small changes by chance. The knowledge of machines can be tracked down to human engineers—i.e., machines are extensions of the humans who create them and lie thus at the practopoietically higher levels of organization (e.g., *top*+1). It took billions of years of biological evolution to create bimetal and arrange it into a thermostat. Thus, the fundaments of the cybernetic knowledge of machines can also be tracked down to biological evolution.

Adaptively more advanced machines i.e., more intelligent machines, should be able to extract their own cybernetic knowledge in high proportion and thus, reduce the role of humans. For example, a thermostat with an additional traverse at the bottom of the hierarchy should be able to extract its own knowledge on how to keep a space comfortably warm. A robot should determine its own behavioral actions to achieve its goals.

## 3. Characterizing systems of different adaptability levels

The central idea of practopoietic theory is that, depending on the number of traverses, there are limitations on how much a system can adapt even if the variety of the system is unlimited. Here we systematically characterize systems of different numbers of traverses, which are labeled as $T_n$, where *n* indicates that number. The most important is the difference in the maximum adaptive capabilities exhibited by systems that have two traverses, as presumed by the current brain theories, in comparison to those that have three traverses and thus, exhibit additional adaptive competencies.

### *3.1 A $T_0$-system: information and structure*

A *$T_0$-system* does not have practopoietically operational capabilities. It exhibits zero traverses and has only one level of (self-)organization. A $T_0$-system is a part or a *structural* component of a larger system. A $T_0$-system can be adapted, but it does not perform any adaptation itself.

Any structural element of a system e.g., a bone in a body is a $T_0$-system, and so is any passive form of information storage, such as a book or DNA. Any tool or instrument, such as a knife, has a maximum of $T_0$-capabilities too. Also, active components e.g., a motor or a computation processor, have $T_0$-capabilities if they are not closing a loop with the environment to which the system adapts.

$T_0$-systems are relevant for practopoiesis as constitutive components of larger, more adaptive systems. They provide support such as *structure* or *information* that is utilized within the system.

Hence, not any object or computation can be labeled $T_0$. To be granted the title, a component must be a functional part of an adaptive system and thus, must already have undergone certain steps of practopoietic organization and knowledge extraction.

### *3.2 A $T_1$-system: control and deduction*

A *$T_1$-system* exhibits one traverse and therefore, involves operations across two levels of organization. This system exhibits minimal adaptive capabilities. Its physical structure enables receiving inputs from the environment and sending outputs.

The cybernetic knowledge of that system may be austere, as in the case of a simple thermostat, or rich as e.g., stored within the connectivity pattern of a large neural network wired-up to input-output devices enabling interactions with an environment. $T_1$-systems can close a loop with the environment in a continuous manner or in a discrete one i.e., acting only when specific conditions are met, for example when a threshold is reached. Hence, in its simplest form, a $T_1$-system can be described as a *control* mechanism, or as a *regulator*. Also, a variety rich $T_1$-system can be seen as an elaborate monitor-and-act machine—a device that responds to events in the environment.

A $T_1$-system can also be understood as a mechanism that extracts knowledge. More formally they can be said to implement *deduction of cybernetic knowledge*: The action for a specific case is deduced (at higher level of organization) from a general rule (at lower level of organization).

In biology, subsystems of an organism can be described as $T_1$ when they perform *homeostatic* functions (Cannon 1932). For example, negative feedback loops for controlling body temperature are $T_1$-systems. The same is the case for the mechanism for regulating blood glucose levels (Ahima & Flier 2000). Reflexes e.g., a stretch reflex (Liddell & Sherrington 1924; Gurfinkel et al. 1974), can also be described as having a single traverse. The rate of gene-expression, which is regulated by a feedback loop, is a $T_1$-system. For example, the excess of tryptophan directly prevents further synthesis of that amino acid (Gollnick et al. 2005). $T_1$-systems are not limited to negative feedback but can implement positive-feedback loops too[16]. Human-made devices can be described, in general, as being limited to $T_1$-capabilities.[17],[18]

The main limitation of $T_1$-systems is excessive variety that would be required to deal with real-life problems. Although such systems can implement in principle any mapping function, in real life this is not enough because the number of combinations of events that an animal or a person could possibly encounter in his/her life in all possible environments that it may live in and in all possible situations that it may encounter, is way too large to be stored in a $T_1$ physical system.[19] Instead, more flexibility is needed to learned selectively only about those environments in which the organisms actually happen to live.

*3.3 A $T_2$-system: supervision and induction*

A *$T_2$-system* consists of two traverses and provides as much a whole new class of flexibility compared to $T_1$, as $T_1$ adds to adaptability in comparison to a $T_0$-system. A $T_2$-system can be understood as granting *supervision* to a $T_1$-system in the form of machinery that monitors the effects that $T_1$ produces on the environment and that has the cybernetic knowledge to adjust the $T_1$ component whenever necessary. The need for adjustment may appear e.g., when properties of the environment change.

A $T_2$-system operates across a total of three levels of organization, the lower traverse relying on the most general form of cybernetic knowledge (the rules of supervision) and extracting knowledge of medium generality (the supervised properties of the system) and then, the higher traverse relying on that knowledge to extract an even more specific form of knowledge (the actual interaction with the surrounding world). Thus, a $T_2$-system can cover more area of the generality-specificity continuum than $T_1$ can (Figure 2B vs. 1C-left).

The additional adaptive capabilities of a $T_2$-system stem from the properties of its middle level of organization. While the cybernetic knowledge at the bottom of the hierarchy is always fixed and the one on the top of the hierarchy changes perpetually with even the slightest change in the environment, the middle level in a $T_2$-system provides a place to store temporary knowledge that may be valid for a while, but which may be changed later if circumstances require so. A $T_2$-system is the first one that is able to learn on its own to control the environment. In other words, while a $T_1$-system controls only the surrounding world, a $T_2$-system controls also itself. Thus, a $T_2$-system can be understood as being capable of *inducing cybernetic knowledge*. It learns how to monitor and act. The process underlying the lower traverse induces the rules that drive the deductions of the higher traverse. For example, a $T_2$-system equipped with a thermometer, a heating pad, a few other components and appropriate learning rules may be able to invent a thermostat and by doing so, extract cybernetic knowledge on how to maintain the environmental temperature constant. In that example, the invention process is the supervisor of the thermostat.

In biology, many examples of $T_2$-supervision can be found. Gene expression mechanisms play the ultimate supervisory role within an organism. The homeostatic function that any organ performs, or the regulation machinery responsible for a reflex, or the feedback loop involved in the response of the immune system—all need to be supervised. Someone has to make sure that they work properly, and make adjustment when necessary. In biological systems, this supervisory role can be traced back to gene expression mechanisms.

Therefore, to keep one variable constant in an unpredictable world, the control mechanism for that variable has to adjust, which means changing some other variables in the system by operations performed by the supervisory systems. In other words, in $T_2$-systems, lower-level traverses have the capability of inducing *allostasis* (Sterling & Eyer 1988; Sterling 2004; Karatsoreos & McEwen 2011): maintaining constancy at one place in the system by making the necessary changes at another place in the system. For example, in a case of dehydration, extensive physiological changes are needed in order to maintain the most critical internal water concentrations in the working range. Urine output is reduced. Veins and the arteries are constricted to maintain blood pressure with less fluid. The tongue and the mouth dry up.

Whereas a minimum of $T_1$-adaptability is needed for Bernard's (1974) milieu intérieur and homeostasis (Cannon 1932), a minimum of $T_2$-adaptive capacities is needed for a system to be able to perform allostasis (Sterling & Eyer 1988; Sterling 2004; Karatsoreos & McEwen 2011). Thus, although allostatic systems are built solely from homeostatic mechanisms (Day 2005), allostasis reflects an increased level of system organization (i.e., $T_2$ is build from $T_1$-components).

### 3.3.1 Neural networks and $T_2$

Supervision i.e., knowledge induction, is also important for organizing neural networks. In the nervous system, plasticity mechanisms play the supervisory role for establishing the anatomy of the system, which in turn determines how the sensory-motor loops operate. Plasticity mechanisms mediate growth of axons and dendrites, formation of synapses and neuronal excitability.

Activity-dependent plasticity is responsible for the development of a nervous system and for its maintenance later (Dubner & Ruda 1992; Ganguly & Poo 2013). A minimum of $T_2$-structure is needed to allostatically change the anatomy (synaptic weights in the mildest form) in order to maintain behavioral functionality of the system as a whole. A recovery after an injury, such as a stroke, also could not occur without a $T_2$-structure and thus, without feedback obtained through exercise. Failure to successfully function at the higher traverse i.e., at the sensory-motor functions of the neural network, induces downward pressure for adjustment by actions of the lower traverse.

While some of the lower-traverse plasticity mechanisms may simply be keeping a neuron within its optimal operational range, such as the up-regulation of excitability following a period of quiescence (Mozzachiodi & Byrne 2010; Hansel et al. 2001; Turrigiano 2012), others may have a more general adaptive function related to the neuron's function in goal-oriented behavior (Buonomano & Merzenich 1998; Draganski et al. 2004; Xu et al. 2009). For example, the reward systems based on dopamine signaling (Wise 1996), can inform a cell whether to make changes in order to produce a more adaptive form of behavior in the future (note that Hebbian learning alone is generally not sufficient to provide an additional traverse[20]).

The higher traverse of a neural system involves de- and hyperpolarization of neural membranes, generation and delivery of action potentials, and synaptic transmission. Here, cybernetic knowledge created by the plasticity mechanisms and stored at the level of anatomical properties of a neuron is used to extract more specific knowledge in the form of the current activity of that neuron. This highest level of organization involves both physiological and behavioral phenomena. Physiological phenomena at that top level are firing rates, inhibition, excitation, neural synchrony, oscillatory activity, etc. Behavioral phenomena are manifested as simple reflexes but also as more elaborated forms of closed sensory-motor loops—such as the willful conscious behavior.

An example of a $T_2$-system that establishes proper connectivity in a network is illustrated in Figure 3. To obtain feedback from the environment at its lowest level of organization, a neuron may monitor the efficiency of its outputs in controlling its own inputs. The presumed rule strengthens connections if the output of a neuron has the power to mute its inputs (Figure 3A) and weakens connections otherwise (Figure 3B). In the most extreme case, the neuron may completely remove a connection defined as un-functional by those rules and seek to create a new one (Figure 3C). These rules grant adaptive capabilities to the system. The rule may be used to establish the network connectivity at first but can also be used later if the environment changes those relationships and the neuron's connectivity needs to be adjusted again. This grants considerable adaptive capabilities to the system. For example, if a reflex is not working efficiently due to muscle fatigue, such a rule can be used to crank up the efficiency of a synapse involved in that reflex—resulting in an improved overall functionality.[21] In man-made devices, $T_2$ allostatic systems are rare and rudimentary[22].

$T_2$-structure helps a system operating in a changing, environment unpredictable at fast time scales in which no preconceived plans can be executed without an occasional need for adjustments to new circumstances, and no rules of behavior can be applied for long time without the need for adapting them according to the altered properties of the world. There is a need for general knowledge on how to adjust in a given situation. The system monitors indicators of successes/failures of the executed actions and modifies its own properties according to more general knowledge of which adjustments should be made, and how.

The limitation of a $T_2$-system is that it can learn efficiently to deal with only one set of behaving rules at the time i.e., with one type of situation. If the system learns to behave in one situation, it has hard time behaving in another situation without extensive relearning thus, finding itself in a need to forget the old knowledge in order to acquire new knowledge. This transition from situation to situation is related to stability plasticity dilemma[23] and is costly both in learning time and in the adaptability that the system can exhibit.

### *3.4 A $T_3$-system: anapoiesis and abduction—the Mind*

To the same degree to which $T_1$ has more adaptability than $T_0$, or $T_2$ than $T_1$, a $T_3$-system has a qualitatively higher level of adaptability than a $T_2$-system. This system can be seen as a higher-order supervisor—or supervision of a supervisor, which, when combined with high variety, gives it unique adaptive capabilities.

The adaptive advantages of a $T_3$-system stem from its expanded capabilities for acquisition of cybernetic knowledge at two different levels of generality. A $T_3$–system has in total four levels of organization, which can be referred to as *top-3*, *top*-2, *top*-1 and *top*. This provides the system with two levels of organization at which it can change its own structure (*top*-1 and *top*-2) i.e., at which it can learn (Figure 4).

The most obvious advantage is more detailed coverage of the generality-specificity continuum of knowledge about the surrounding world. However, there is a qualitative leap in the adaptability that comes from this additional traverse when variety is high at each level of organization. The system can juggle much knowledge internally from a general to a specific level and back: With a minimal hint from the environment on what is about to come, a previously acquired knowledge about the upcoming activities can be pulled out from the general level and poietically instilled at a more specific level.

As a consequence, a $T_3$–system is not only capable of learning how to control but it can also learn how to learn quickly. The mentioned slow adaptation process may turn into a process as quick as what it takes to recognize a pattern. This is made possible by the intermediate traverse out of the total of three traverses that the system possesses (Figure 4A). This is the traverse in a sandwich i.e., it is the one whose both ends meet other traverses: Its lower-end knowledge is not fixed but can be changed; Its higher end-knowledge is not an output but is still a part of the system. The consequence is that this middle traverse can give the system unprecedented level of adaptability, which, with sufficient variety, leads to nothing short of the ability to think.

The middle traverse can be understood as *reconstructing knowledge* at *top*-1 that has been extracted once but lost since. In $T_3$-systems, the knowledge at *top*-1 can be treated as temporary, while more permanent version is stored at *top*-2, which has also a more general (abstract) form. This generalized knowledge is stored by the learning rules at *top*-3 (Figure 4B). Then, when needed, *top*-1 knowledge can be reconstructed from *top*-2 by a relatively brief interaction with the environment. Ultimately, it is *top*-1 that controls behavior directly (*top* is the actual behavior). It is the level of organization that embraces all of the following: the current position of limbs, the current tension in muscles, and the current depolarization of neural membranes. And it is *top*-1 that determines how the depolarization of membranes change will next and thus, how will the tension in muscles and limb positions change. However, it is the general knowledge at *top*-2 that does the control in the background because cybernetic knowledge at that level enables flexible exchange of the contents at *top*-1. Thus, with each change in the general properties of the environment, the system may not need to relearn everything from scratch and extract cybernetic knowledge from the environment again, as a $T_2$-system would need to do. Instead, given that the traces of previous encounters with similar situations have been stored at two levels below, the knowledge can be brought back up now quickly.[24]

That way, a familiar situation i.e., a set of environmental properties, needs to be detected to initiate reconstruction, but the details associated with that situation need not be learned all over again. Many of the details are already pre-stored and can be easily "pulled out" in a given situation or context. Thus, a $T_3$-system can also be understood as implementing *situation-dependent* or *context-dependent supervision*.

This reconstructive traverse from *top*-2 to *top*-1 organization level is referred to as *anapoiesis*, from Ancient Greek ανά (ana) meaning "over, again". The term refers to the repeated creation of knowledge through reconstruction from a general depository to a more specific form.

Anapoiesis is an additional intermediate generator of variety at *top*-1. It is triggered whenever the environment significantly changes and downward pressure for adjustment is exerted onto the monitor-and-act units at the level *top*-2. If no significant pressure has been exerted at *top*-3 and if the system eventually succeeds in removing the adjustment pressure by relying on *top*-2 / *top*-1 only, then a relatively easy solution to the problem has been reached. The system has successfully reconstructed knowledge from its past experiences and used anapoietic reconstruction to guide its behavior in a given situation.

In contrast, if the downward pressure for adjustment reaches all the way to the bottom and thus, the monitor-and-act units at the level *top*-3 are informed of a need to make changes, anapoiesis alone has likely not been sufficient to satisfy the needs of the system. A new, unfamiliar situation is encountered! In that case, a $T_3$-system adapts by deploying its unique capability to make changes to the general knowledge driving anapoiesis—creating new knowledge at the level *top*-2 for a new type of situation. Thus, the full dynamics of the practopoietic cycle of causation in a $T_3$-system includes anapoiesis as the middle traverse (from organization level *top*-2 to *top*-1) but also the verification process, which necessarily engages the top traverse (from *top*-1 to *top*), and the adjustment of the general knowledge (from *top* to *top*-2), which is engaged whenever anapoiesis fails. This describes the full global workspace (Baars 2005) of a $T_3$-system.

In artificial neural networks, anapoiesis may provide a general solution to the problem of the stability-plasticity dilemma, as it enables dealing with both general and specific knowledge[25]. But anapoiesis needs not only underpin cognitive-like operations. An example of is the creation of phenotype from genotype[26]. Here, one has to note that if individuals are $T_3$-systems, an entire evolving species of such individual forms a $T_4$-system. Systems bigger than $T_4$ do not seem to be necessary for survival on the planet earth[27].

*An example: Applying rules vs. learning them.* An adaptive system changes the rules of behavior given a change in the situation. In a $T_3$-system this is done at two different levels: At one level the already known, previously used, rules are being reactivated. This level employs anapoiesis. At the other, lower level novel rules are being extracted.

A toy example of rule reactivation and extraction is a *Wisconsin card-sorting test* used in clinical assessment of executive functions (Berg 1948). In this test a participant sorts cards from a deck to match one of the properties of the reference cards placed in front of the participant. The deck and the reference cars should be matched either in color, shape or the number of items. Importantly, the participant is never told explicitly which property needs to be matched and is only given feedback in form of “correct” or “wrong”. In addition, the sorting rules unexpectedly change during the task and the only indicator of a change is the feedback “wrong”. The participant’s task is to find out the new rule.

This test applies matching rules that are intuitive (e.g., matching red color with another red color) and thus, in a way, already known to the participant. For that reason, the problem can be understood as engaging a $T_2$-system: An anapoiesis-like process activates one of the rules at the time (at *top*-1) from the repertoire of the known rules (stored at *top*-2). Thus, no induction of novel rules is necessary and hence, no traverse that operates below *top*-2 seems to be involved.

Neural networks implementing such pre-existing rules have been created and demonstrated to mimic human performance (Levine & Prueitt 1989; Dehaene & Changeux 1991; Parks et al. 1992; Carter 2000; Kaplan et al. 2006). In these systems, the *top*-2 knowledge has been either hand-coded (e.g., Dehaene & Changeux 1991) or pre-trained (e.g., Carter 2000)—in either case acquisition of those rules requiring an intervention from a side of a human programmer. Here, human programmers provided the required deeper levels of adaptability that $T_2$-systems could not possess.

However, one can envision an extended version of Wisconsin card-sorting test that requires the system to extract those rules because they are not intuitive and already learned. There are many rules possible that are not so intuitive. For example, the green reference may have to be matched to a red deck card; or the count of three items to a square shape, etc.

So, what if the test is made suddenly in such a way to be more difficult and not to rely on the most intuitive types of associations? The number of potential rules becomes too large to be hand-coded or pre-wired—thus, more like a real-life situation. The learning time becomes longer and a number of errors becomes larger. In that case a full $T_3$-system is required to explore the space of possibilities—largely by trial and error—and eventually acquire new *top*-2 knowledge. The system has to use the feedback to make changes at *top*-2 level iteratively. But upon successful learning, the system can operate again quickly through anapoietic reconstruction from *top*-2 to *top*-1.

### 3.4.1 Peristasis

The adaptive capabilities of a $T_3$-system can also be understood from the perspective of regulating system variables key for survival, or homeostasis (Cannon 1932; Bernard 1974), and the distinction between homeostasis and allostasis (Sterling & Eyer 1988; Sterling 2004; Karatsoreos & McEwen 2011). A $T_3$-system has adaptive capabilities that exceed those of allostasis. The bottom traverse can adjust a $T_3$-system to its habitat such that it can perform allostasis more efficiently—i.e., fast reconstruction means less allostatic load (McEwen & Stellar 1993). If the organism is exposed to extreme allostatic pressure e.g., cold, a $T_3$-system is able to adjust such that allostatic pressure is reduced, by e.g. growing fur, and thus a smoother physiological operation is ensured: The knowledge at *top*-3 is used to reconstruct properties of the system at a higher level of organization. This adjustment reflects a more elaborate form of adaptation to the surrounding world than allostasis i.e., a higher level of organization, and can be referred to as *peristasis*, from Ancient Greek word περί (peri) meaning “around”. Peristasis refers to “staying stable by understanding (or grasping) the conditions for adaptation that apply to the current situation”. In our extended, more difficult version of Wisconsin card-sorting test, peristasis would be achieved by the acquisition of a new set of rules. That way, by activating one of them, a $T_3$-system keeping stable the most important variable of that task: the feedback “correct”.

### 3.4.2 Logical abduction

$T_1$- and $T_2$-systems perform cybernetic operations that correspond, respectively, to logical deduction and induction. There is an advanced form of logical-deduction that can be performed with its full power only by a $T_3$-system. Anapoiesis of a $T_3$-system can be described as a use of past knowledge to guess which knowledge is correct for the given situation and then evaluating the degree to which the guess matches reality, and adjusting the discrepancies that may appear. The corresponding guess-based logical operation is known as *abduction*, introduced to account for the inferences made on the basis of the best hypothesis given the available knowledge (Peirce 1903). Abduction involves validation and correction of the guess, which requires iteration of the abducing steps. In principle, a $T_2$-system is sufficient to implement logical abduction provided that a large-enough knowledge-base exists at *top*-2 level from which the hypotheses can be drawn. The process of establishing and refining this knowledge-base requires one more traverse i.e., a $T_3$–system. Only A $T_3$-system can make a guess through the anapoiesis of the existing knowledge, validate the guess by interacting further with the environment, and then adjust the knowledge-base as necessary: If the guess turns incorrect, engagement of *top*-3 monitor-and-act units is needed. This results in equipping the system with new knowledge for abductions, which means that learning takes place by applying the lowest of the three traverses. For example, in our extended Wisconsin card-sorting test, the simple intuitive rules may be abduced first, but then rejected in light of the feedback. Another rule may be abduced next, tested, rejected, etc. In a probabilistic form, abduction is described by Bayes' theorem, which has been argued to be relevant for brain operations (Friston 2010; Shipp et al. 2013; Friston et al. 2012; Clark 2013). However, Bayes' theorem cannot provide a complete description of $T_3$-systems and hence, of the mind because there is nothing in Bayes' approach that would correspond to the lowest traverse of the three—learning probabilistic priors. Thus, Bayesian inferences enter practopoietic systems through the top two traverses and hence, thorough anapoiesis. The present theory offers a general approach through $T_3$-systems, covering also the learning of the knowledge on the basis of which inferences are made.

## 4. Discussion

Much of the biological knowledge and skills can be stored in a form of cybernetic variety, but it is only the levels of (self-)organization that bring about the capacity to acquire knowledge. Practopoietic theory proposes that these levels of organization are achieved through traverses: Knowledge acquisition proceds always from general to specific.

The implication is that, to achieve intelligence, a system needs not only variety in a form of e.g., network connectivity, hardware components, if-then statements, etc., but also a feedback from environment though which the variety is being adjusted. The key contribution of practopoietic theory is the generalization of the role of feedback: In any given system, the principles by which the variety is adjusted can be also adjusted themselves by yet another set of principles, and so on. And each set of principles can have its own variety. This generalization results in a hierarchy that can in principle grow indefinitely. Each step in this hierarchy is one traverse of cybernetic knowledge.

It follows that variety and traverses should be considered as somewhat orthogonal in contributing towards the total intelligence of the system: Variety is about knowing what to do; Traverses are about acquiring this knowledge. Both components are essential and neither alone can provide powerful intelligence. Thus, no matter how much variety one may add to a system e.g., in a form of neurons and connections, the system may still not be able to produce human-like mental capabilities if it does not have enough traverses.

Practopoietic theory allows us to analyze the adaptive competences of systems with differing numbers of traverses. These competences range from simple information storage at $T_0$ (no traverses) and deductions from this information at $T_1$, up to induction of cybernetic knowledge at $T_2$, and learning to perform abduction at $T_3$ (i.e., three traverses). The properties of different systems are summarized in Table 1.

A conclusion is that a $T_1$-system cannot possibly have enough variety to deal with the combinatorial explosion of the real-life situations of a human person. A $T_2$-system does not solve this problem satisfactorily either, as it requires forgetting old knowledge when learning new one. But a $T_3$-system appears to have enough flexibility to deal with the richness of a real life. This system can change itself on two levels: it can learn abstract rules and reconstruct from them concrete ones in a particular situation. A $T_3$-system takes also advantage of the fact that with more adaptability levels the system can be smaller in total size and yet, produce the same or higher amount of variety than systems with fewer such levels.

Further analysis of the unique properties of $T_3$-systems indicated that what is particularly missing in our brain theories, and also in our technology of AI algorithms (Kurzweil 2005), is the middle traverse of those systems—referred to as anapoiesis. Thus, to address the mind-body problem successfully (Descartes 1983/1644; Popper 1999; Chalmers 1999; Rust 2009), practopoietic theory suggests that it is necessary to consider $T_3$-systems with variety large enough to foster powerful anapoiesis.

Much of what we know about human cognition supports the idea that our minds are $T_3$–machines relying on anapoiesis:

*4.1. Reconstructive memory*: There is evidence that recall from human memory is reconstructive by its nature (Schacter et al. 2000; Squire 1992; Burgess 1996), and that working memory capacity is directly determined by reconstructive capabilities by a process known as chunking (Miller 1956; Cowan 2001)[28]. Thus, both of these phenomena may fundamentally rely on anapoietic reconstruction from general to specific knowledge. Similarly, past stimulation builds expectancies for later stimulus processing (Albright & Stoner 2002; Nikolić 2010). These context-induced expectancies possibly require anapoietic processes too: Expectations may be produced by adjustments at the level *top*-1 and using the knowledge acquired previously at *top*-2.

Efficient management of expectancies is adaptive. As an animal is behaving, it needs to activate different situational knowledge on momentary basis. Every new situation that it enters requires different knowledge on what can be expected to happen and what may need to be done. For example, as a hedgehog leaves shelter, enters open space, moves into woods, detects food, etc., each situation implies different expectancies. These situations can exchange literally every few steps of a walk. It is more efficient to reactivate existing knowledge in a form of working memory contents and expectancies than to re-learn it from scratch.

*4.2. Downward pressure for adjustment*: Evidence indicates that the degree to which these working memory and expectancy mechanisms are engaged depends on downward pressure for adjustment. Slow, capacity-limited working memory resources, or controlled processes, are engaged typically when difficulties arise using quick and capacity-ample automatic processes (Shiffrin & Schneider 1977; Stanowich & West 2000; Kahneman 2003, 2011). This suggests that the strength of the downward pressure for adjustment plays a role in activating anapoietic mechanisms. This pressure is particularly extensive when the organism encounters novel situations to which it yet has to find suitable knowledge at *top*-1.[29]

Similarly, evidence indicates that explicit long-term memories are facilitated by downward pressure for adjustment exerted from the contents of working-memory. Memories for verbal materials are good when their relation to a certain context is processed i.e., when the contents are crunched intensively by working memory. In contrast, the memory is poor when only sensory aspects are processed (Craik & Lockhart 1972). Similarly, in vision, long-term memory for visual patterns improves linearly with the time over which the patterns are processed in visual working-memory (Nikolić and Singer 2007). These results can be interpreted as downward pressure for adjustment exerted by anapoietic operations on *top*-3 mechanisms to form novel long-term memory at *top*-2 level.

*4.3. Concepts*: The capability of the human mind to conceptualize the world (Barsalou et al. 2003; Gallese & Lakoff 2005) may be accounted for by anapoiesis of knowledge too. Our conceptual knowledge, stored in long-term memory, consists of generalized, abstract rules of interacting with the world (e.g., Barsalou et al. 2003). Hence, to apply this knowledge to a specific case, there is always a need for a matching operation: general principles should be matched to a specific situation. This is where anapoiesis comes to aid: When an object is encountered, it may be categorized by matching the generalized knowledge at *top*-2 to the sensory inputs coming from *top*. The result is the knowledge constructed at *top*-1 that is specific to that object. Only then can the system interact with that object successfully. For example, thanks to the anapoiesis of concepts we may be able to drive a car and avoid collisions in novel traffic situations that never occurred before: General driving rules are applied to each specific situation.

Because it stores all our concepts, the level *top*-2 can be referred to as *ideatheca* (Greek for "place where concepts are stored"). These concepts in ideatheca should be shielded from the current activity at *top*-1, and should be adjusted only by the operations of monitor-and-act units at *top*-3.

*4.4. Ambiguities and problem solving*: Biological minds are distinguished from machines largely for their ability to resolve ambiguities (e.g., Kleinschmidt et al. 1998) and cognitive problems in general (Sternberg

& Davidson 1995; Jung-Beeman et al. 2004). Anapoietic reconstruction may be the key behind those intellectual capabilities. Natural to a $T_3$-system is a reiteration of anapoiesis in case that the first round was not successful in removing the pressure for adjustment. In case of failure, the pressure remains and thus, the need to continue with anapoiesis remains too. With each subsequent anapoietic iteration chances to find a solution may improve due to the work done by the preceding anapoietic steps. Although they failed, they may have brought the system closer to the solution than it was before. An important part of that is the adjustment pressure that is, due to the failures of anapoiesis, exerted on the lower level i.e., long-term memory.

This dynamics of failure and pressure may underlie the process of abduction. For example, in an ambiguous situation (Is this a predator or a prey? A friend or a foe?), a $T_3$-system may first abduce a hypothesis, and by doing so, drive the actions of the sensory-motor system towards obtaining further sensory inputs to test that hypothesis (e.g., by directing gaze). The hypothesis may be then confirmed or rejected. If rejected, abduction of a new hypothesis may require concurrent changes at *top*-2 consistent with the knowledge that the first hypothesis was incorrect. The process may then continue. This iterative dynamics of resolving ambiguities, from *top* to *top*-2 (i.e., from behavior to ideatheca), can eventually produce appropriate cybernetic knowledge at level *top*-1 that is original and different from anything in the past to a sufficient degree to be qualified as an insight or a creative solution to a problem.

*4.5. Anapoietic cognition*: In general, a property of the intermediate anapoietic traverse, lying in the sandwich between sensory-motor loops and plasticity, is that it allows for reorganization of knowledge without immediately executing behavior. That is, anapoiesis may not act immediately towards the main goal—i.e., towards resolving the main downward pressure for adjustment. Instead anapoiesis may act first towards sub-goals—postponing the main behavioral actions until the conditions for actions are ready. These sub-goals may involve behaviorally covert operations, which, when becoming elaborate, may manifest themselves as cognition.

Thus, we may hypothesize more generally that our entire cognition is based on anapoiesis: An arrival at a Gestalt of a percept (Köhler 1929), attention successfully directed (Treisman 1980; Posner & Petersen 1990),[30] stimulus recognized (Furmanski & Engel 2000), object mentally rotated (Kosslyn et al. 1998), a logical conclusion inferred (Clark 1969), a decision reached (Bellman & Zadeh 1970), a problem solved (Sternberg & Davidson 1995; Jung-Beeman et al. 2004)—may all be end-results of anapoiesis. In cognitive science, the outcomes of these activities are operationalized as working memory contents, focus of attention, recall, imagination, expectancies, biases, accumulation of evidence, etc. In practopoietic theory, these resulting mental contents can collectively be referred to as cybernetic knowledge at *top*-1 level activated from ideatheca.

*4.6. Awareness*: Anapoietic process may also account for the capability of biological systems to be aware of the surrounding world. Anapoiesis, has never a full internal "peace" of uninterrupted operation like e.g., a computer algorithm would have when factoring a large number (it is not boxed). Instead, anapoietic process is continually bombarded by downward pressure for adjustment as a result of an unceasing influx of sensory inputs. Anapoietic process has to integrate all the inputs through its equi-level interactions, and this results in a form of continuous peristasis—i.e., perpetually adjusted knowledge of what is currently out there in the surroundings, even if it is irrelevant for the current task.

That way the systems satisfies Ashby's good regulator theorem for the current environment. The great adaptive advantage is that this knowledge can be used immediately if the distractor becomes suddenly relevant for the task, or relevant in any other way. For example, while hunting, an animal may have to integrate irrelevant auditory inputs such as the sounds of a water stream. But this very integration enables detecting effectively changes in that sound, which may then be essential for survival as they may indicate e.g., the presence of a predator. Thus, eventually, the inability to switch off and the necessity to integrate may lead to particularly adaptive behavior: Stopping the hunting and seeking shelter. Thus, due to the equi-level interactions across the monitor-and-act units of anapoiesis the knowledge is organized at the level *top*-1 such as to take into account everything that enters through senses, not only the information related to the current task. That way, the system becomes aware of its surrounding world.

*4.7. AI and understanding*: The difference between $T_3$- and $T_1$-systems may be the difference between what Searle (1980; 2009) referred to as understanding on one hand, and the input-output mapping programmed into computer algorithms on the other hand. Searle distinguishes between semantics-based processing by human mind and syntax-based manipulation of symbols by machines. In his Chinese Room argument, Searle illustrates that computer algorithms, that are being programmed and thus provided the knowledge from the outside, cannot understand what they are doing. Hence, such algorithms cannot think, and thus cannot provide human-like artificial intelligence (or strong AI).[31] Practopoietic theory explains what these algorithms are missing: An algorithm that operates through syntax is a $T_1$-system, while understanding with all the conceptualizations requires at least a $T_2$-system with a rich ideatheca: Understanding comes through the very nature of the transcendence of knowledge in practopoietic system. The activation of a concept from ideatheca into *top*-1 is in the same time an abductive inference made on "symbols" and a symbol grounding process[2]. Thus, to infer anapoietically is to simultaneously verify through feedback both the premises and the conclusion of the inference. This in turn means that for full mental functioning a $T_3$-system is needed, as rich ideatheca can be established only if monitor-and-act units exist at level *top*-3. The existing artificial intelligence systems that have $T_2$-structure do not have any capability to enrich or improve their ideatheca. Thus, only the human programmer is the one who understands, while a $T_1$-machine executes that code without any understanding.

Hence, practopoietic theory prescribes that strong AI can be created only with multi-level interactions with its environment based on a $T_3$-system architecture. This requires the system to acquire general knowledge stored at ideatheca by its own interactions with the environment i.e., without being hard coded by a programmer. The system begins to understand and posses semantics when, during its operations, bits of knowledge are used continuously from ideatheca to set anapoietically the properties of the level *top*-1. It is the continuous guessing (abduction) of this process and the continuous corrections of those guesses that make the system effectively conscious.

The challenge for AI is then to endow artificial systems with the needed seed-knowledge i.e., the most general learning mechanisms at the level *top*-3, that are suitable for acquiring knowledge in ideatheca. Only then can the system achieve mind-like operations that activate a more specific form of that knowledge at *top*-1, which then in turn enables the system to exhibit intelligent overt adaptive behavior at *top*.

*4.8. The problem of downward causation.* A question of how can mind can affect body is one of the central problems in the mind-body problem (Sperry 1969, 1980; Campbell 1990; Bedau 2002; Bateson 2004; Robinson 2005; Thompson 2007; Noble 2008a, 2008b)? This interaction mental-to-physical is often described as the problem of *downward causation*, and is an issue that has been difficult to resolve satisfactorily. On one hand, from the perspective of a biologist or psychologist it seems obvious that downward causation takes place: Mental conflicts cause physical symptoms; Thoughts cause behavior; Behavior causes expression of genes (Freud 1915; Sperry 1969, 1970, 1980, 1993; Noble 2008a, 2008b). However, when dissected formally within the framework of the theory of emergence (Meehl and Sellars 1956; O'Connor 1994), the concept of downward causation seems to defy logic. Some properties P of a system are said to emerge from other properties O of the system if, and only if, P supervenes on O—i.e., if the whole supervenes on its parts. For example, the behavior of a dynamical system supervenes on the equations describing that system. Consequently, system's behavior emerges upwards from the properties of those equations. The problem arises when one attempts to define downward causation within such systems in which O supervenes on P. The concept of downward causation seems not to bring any new explanation to the dynamics of the system that has not been already accounted for by upward causation. Downward causation becomes hence "inconsistent" (Szentagothai 1984), "mysterious" (Bedau 1997, 2002) and "can at best supply a false sense of satisfaction" (Robinson 2005).

Practopoietic theory offers a way to reconcile these opposing views from intuition about biological systems on one hand and the strict theoretical analysis of system dynamics on the other hand. The key novelty that practopoiesis offers to system theory is the idea that one should *not* attempt to understand adaptive systems as supervenient relations between its P's and O's. Instead, the system should be decomposed into poietic relations, which by their very nature are not supervenient. Let us take the example of proteins and genes: Physiological functions of a protein supervene on its own atoms and their spatial arrangement. However, physiological functions of a protein do not supervene on the atoms underlying the DNA code for that protein, nor does it supervene on the atoms of the ribosome that synthetized the protein. Instead, the relation to those gene expression mechanisms is the one of creation (poiesis). Practopoiesis focuses on those poietic relations that exist among the components of the system.

By the uni-directionality of a poetic process, practopoietic systems necessarily have higher and lower levels of organization and it is the interactions among those levels (components, monitor-and-act units) that enables us to provide a consistent, non-mysterious explanation of interactions that involves effects towards up and down. In other words, although the system as a whole is autopoietic, its components mutually have allopoietic relationships (*allo* stands for "other"). It is these allopoietic relationships that are central for understanding adaptive behavior.

Consequently, the emergence of mind and behavior is not primarily a question of the relationship between the elementary parts and the whole, but between components at intermediate levels of aggregation, which represent neither the wholeness nor the most elementary parts. Poietic approach to self-organization and emergence offers a very different perspective than does supervenient emergence.

In practopoietic systems, interactions with the environment account for much of the causation towards down. The concept of trickle-down information explains other types of downward effects. Effects towards up are poietic and require knowledge shielding. Most notably, the theory proposes existence of dedicated physical structures responsible for storage of the knowledge at each level (e.g., ideatheca). As a result, somewhat counter intuitively, the mind affect behavior towards *up*—as the mind lays at one step lower level of organization than the actual behavior.

### *4.9. Empirical predictions*

These present insights into the functions of the three traverses can be used to guide formulation of testable hypotheses about the physiological underpinnings of an anapoietic brain. The three most important empirical predictions are:

Prediction 1: Cognitive operations are implemented chiefly through neural adaptation

The mechanisms of anapoiesis should have the following properties: First, they should operate on a time scales slower than spiking activity but, in the same time, they should be faster than plasticity. The time-scale should correspond to the pace with which our mental operations occur: recognition, recall, decision, mental imagery, and others. Studies on response times indicate that the underlying physiological mechanism should not be faster than about 100 ms and probably not slower than a second or two. Second, the underlying mechanisms should be local i.e., they should not be implemented through distributed network operations. The networks already serve the top-most traverse by mediating sensory-motor loops. Thus, we should look for an implementation separated from the network mechanisms. Finally, the mechanisms of anapoiesis should effectively change the properties of these networks.

One particular physiological mechanism seems to fit these requirements. This mechanism is known as *neural adaptation*. If a stimulus is presented continuously, neurons typically exhibit vigorous response only initially (high firing rates known as on-response), and very soon, often already after 100-200 ms in the cortex, the response reduces to much lower levels. This phenomenon is known as neural adaptation. In cortex, neural adaptation has the timing consistent within the timing of cognitive operations (e.g., Nikolić and Gronlund 2002; Nikolić and Singer 2007). Also, neural adaptation takes place locally, within a cell.

Every sensory modality seems to exhibit such adaptation: vision (Hurley 2002), audition (Jerger 1957; Ylikoski & Lehtosalo 1985), touch (Jones, Gettys, & Touchstone 1974), and olfaction (Dalton 2000). But the most important reason to consider neural adaptation as the underpinning of anapoiesis is the fact that adaptation does not result from fatigue of neurons (as has been often incorrectly presumed) but that neural adaptation rather reflects a certain form of cybernetic knowledge—i.e., the neuron has made a "decision" not to fire. This is easily seen in the stimulus to which the neuron has adapted is removed (the so-called off-response). Fatigue hypothesis would predict a further reduction in neural activity (a combination of a reduced drive into the cell and the fatigue). What happens is that off-responses are typically as vigorous as the initial on-response (e.g., Nikolić et al., 2009). Importantly, off-responses are also more differentiated i.e., more unique for each particular stimulus than are the on-responses (Nikolić et al., 2009).

Neural adaptation shapes the response properties of the network. Responses to subsequent stimuli depend strongly on the properties of the stimuli to which the cells have adapted (e.g., Eriksson et al. 2010, 2012). Thus during adaptation, while becoming less sensitive to one type of inputs the system becomes more sensitive to other inputs.

These data are consistent with the function of the middle traverse in a $T_3$-system. Neural adaptation is thus a mechanism that can provide sufficient flexibility (network changes) and context sensitivity (reflect the properties of the current situation) to quickly re-route the flow of information to thus, to produce novel sensory-motor mappings in the network. The mechanisms of neural adaptation are hence in the position to implement anapoietic machinery. The implication is that any current pattern of adaptation across neurons (i.e., neurons that are, and are not adapted at any given time) constitute the neural correlates of: recognition of perceived objects, directed attention, contents of working memory, decisions made, etc. In contrast, properties of anapoiesis are inconsistent with the idea of being implemented by dedicated brain areas[32].

Prediction 2: The mechanisms of neural adaptation learn

If the above hypothesis is correct and neural adaptation is one of the key mechanisms responsible for anapoietic processes in the mind, the following prediction necessarily follows: The rules by which neurons adapt are not fixed but are being learned. As we acquire new knowledge i.e., enrich ideatheca, we learn when to adapt our neurons (and when not)—the properties of adaptation mechanisms depending on the history of interactions with similar stimuli. This in turn means that the properties of neural adaptation mechanisms can be altered in by appropriate experimental manipulation.

Prediction 3: Ideatheca with shielded knowledge

The physiological underpinning of ideatheca is unlikely to be the synapse. First, by their nature, synapses are an integral part of the *top*-1 level, and hence, cannot store *top*-2 knowledge in the same time. Second, synapses do not seem to have the stability required to store life-long memories characteristic of ideatheca (e.g., Holtmaat and Svoboda 2009). Therefore, yet another surprising prediction that can be derived from a tri-traversal theory of mind: Our semantic knowledge, such as skills and declarative facts, is not primarily stored in synapses i.e., in the network architecture (although this architecture needs to be fine adjusted too). Rather, this knowledge that sits available to the mind even if not being activated in a while (sometimes in years) is stored by some other means—those that determine when and how a cell will adapt its responses. These, currently unknown structures should have the following property: They should be shielded from the activity of the network such that the network cannot directly alter these structures. These structures can be altered only by specialized monitor-and-act units that have very general knowledge about when such learning should be made. These most-general monitor-and-act units should lay at *top*-3 level and hence, should be driven by gene-expression mechanisms.

*4.10. What is thought?* Finally, practopoietic theory allows us to address the important question on what the nature of a thought is (e.g., Baum 2004). The most fundamental difference between an anapoiesis-capable $T_3$-system and the classical approach based on $T_2$-systems is in the number of traverses that need to be engaged during a cognitive act. Given that in each of the two systems the lowest traverse provides learning, which is normally slower than the act of thinking, only the traverses that remain can serve the mental operations. This means that the classical approach allows for only one traverse while an anapoietic system has two. Thus, a classical system uses a single traverse to implement both the mental operations and generation of behavior. A $T_3$-system separates those in two traverses, one on top for sensory-motor loops and another for anapoiesis. This implies a fundamentally different nature of the thought process presumed by the two approaches. Rather than relying on the global dynamics of a neural network by the classical approach, local adjustments of network components through neural adaptation produces mental operations in a $T_3$-system. Thus, the present theory suggests the following: *A thought is an adaptation*.

The function of the network is then to execute the sensory-motor loops and thus, to test each newly created adaptation pattern by collecting feedback. This anapoietic approach is consistent with the following properties of the mind: i) reliance on interactions with its surroundings (e.g., 4EA-cognition: embodied, embedded, enactive, extended, and affective)(e.g., Gibson 1977; Brooks 1991; Noë 2012); ii) logical abduction (Peirce, 1903), and iii) semantic understanding as opposed to syntactical symbol manipulation characteristic of machines (Searle 1980).

In addition, adaptation can be contrasted to computation: A computation: i) turns sensory inputs into states that represent properties of the outside world (symbols); ii) is most effective if the operations of inference are isolated from the outside world (boxed); iii) does not bring new information about the outside world besides that already present in the input symbols (deduction); iv) the meaning is assigned to states i.e., symbols are grounded[2], by a process separate from the inference machinery that manipulates those symbols (syntactic); and v) for the same inputs reliably produces the same outputs (faithful). In contrast, an adaptive thought combines computation iteratively and is guided by feedback to produce something more substantial and more alive: i) Adaptation mechanisms take as input cybernetic knowledge at one level of organization (monitor-and-act units) and extract as output new cybernetic knowledge at another level of organization (other monitor-and-act units); ii) This extraction of new knowledge can only be effective if the system is not closed but integrates continual feedback from the outside (permeable); iii) In anapoietic systems new knowledge is extracted through an iterative sequence of guesses and tests (abduction); iv) Operations necessary for symbol grounding are implemented by the same processes that produce abductive inference—i.e., the process of inference is indistinguishable from the process of extracting meaning (semantics); v) The outputs of adaptive systems depend considerably on the feedback received from the environment, opening possibilities for producing something new (creative). As a result, the processes underlying the emergence of computation can only be described as supervening, whereas adaptation can be understood also as a self-organized process of non-supervening poietic relations. The properties of thought presumed by classical and anapoietic systems are summarized in Table 2.

## 5. Outstanding questions for future studies

1) Do mechanisms exist in the brain that enable the process of neural adaptation to learn?

2) Can a $T_3$-system with a large ideatheca provide a foundation for a general theory of psychology?

3) How can practopoiesis guide development of artificial intelligence?

4) Can the principles of practopoietic hierarchy help us better define life and understand its origins?

5) Does practopoietic theory suggest answers for most difficult questions in philosophy of mind and epistemology?

## 6. Conclusions

In conclusion, adaptive intelligence requires not only massive storage of knowledge in a form of a rich network architecture but also a sufficient number of adaptive levels to acquire, adjust and manipulate that knowledge. Our brain theories, empirical investigations, and AI algorithms should consider systems with three such adaptive levels ($T_3$- systems) capable of producing anapoiesis and having a large ideatheca.

**Acknowledgments**

The author would like to thank Shan Yu, Matej Pavlić, Viola Priesemann, Raul C. Muresan, David Eriksson and Kleopatra Kouroupaki for valuable discussions and comments and Wolf Singer for support. This work was supported by Hertie-Stiftung and Deutsche Forschungsgemeinschaft.

## References

Abraham, W. C. & Bear, M. F. (1996) Metaplasticity: the plasticity of synaptic plasticity. Trends in Neurosciences 19(4): 126-130.

Ahima, R. S. & Flier, J. S. (2000) Adipose tissue as an endocrine organ. Trends in Endocrinology & Metabolism 11(8): 327-332.

Albright, T. D. & Stoner, G. R. (2002) Contextual influences on visual processing. Annual Review of Neuroscience 25(1): 339-379.

Alvarez, G. A. & Cavanagh, P. (2004) The capacity of visual short-term memory is set both by visual information load and by number of objects. Psychological Science 15(2): 106-111.

Anrew, A. M. (1979) Autopoiesis and self-organization. Cybernetics and System 9(4): 359-367.

Ashby, W. R. (1947) Principles of the self-organizing dynamic system. Journal of General Psychology 37: 125–128.

Ashby, W.R. (1956) An Introduction to Cybernetics. Chapman & Hall, 1956, ISBN 0-416-68300-2

Awh, E. & Jonides, J. (2001) Overlapping mechanisms of attention and spatial working memory. Trends in Cognitive Sciences 5(3): 119-126.

Baars, B. J. (2005) Global workspace theory of consciousness: toward a cognitive neuroscience of human experience. Progress in Brain Research 150: 45-53.

Bechtel, W., & Richardson, R. C. (1993). Discovering complexity. Princeton, NJ: Princeton UP.

Baddeley, A. D. (1993) Working memory or working attention? Clarendon Press/Oxford University Press.

Barsalou, L. W., Kyle Simmons, W., Barbey, A. K. & Wilson, C. D. (2003) Grounding conceptual knowledge in modality-specific systems. Trends in Cognitive Sciences 7(2): 84-91.

Bateson, P. (2004) The active role of behaviour in evolution. Biology and Philosophy 19(2): 283-298.

Baum, E. B. (2004) What is thought? The MIT Press.

Beadau, M. (1997) Weak emergence. In J. Tomberlin, ed., Philosophical Perspectives: Mind, Causation, and World. 11: 375-399.

Bedau, M. (2002) Downward causation and the autonomy of weak emergence. Principia: An International Journal of Epistemology 6(1): 5-50.

Beer, S. (1974) Designing Freedom. CBC Learning Systems, Toronto; and John Wiley, London and New York.

Beer, S. (1979) The Heart of Enterprise. John Wiley, London and New York. Reprinted with corrections 1988.

Bellman, R. E. & Zadeh, L. A. (1970) Decision-making in a fuzzy environment. Management Science 17(4): B-141.

Berg, E. A. (1948) A simple objective technique for measuring flexibility in thinking. The Journal of General Psychology 39(1): 15-22.

Bernard, C. (1974) Lectures on the phenomena common to animals and plants. Trans Hoff HE, Guillemin R, Guillemin L, Springfield (IL): Charles C Thomas ISBN 978-0-398-02857-2.

Bernstein, N. (1967) Coordination and regulation of movements. New York: Pergamon Press.

Bi, G. Q. & Poo, M. M. (1998). Synaptic modifications in cultured hippocampal neurons: dependence on spike timing, synaptic strength, and postsynaptic cell type. The Journal of Neuroscience 18(24): 10464-10472.

Brooks, R. A. (1991) Intelligence Without Representations. Artificial Intelligence Journal 47: 139-159.

Brooks, R. A. (1999) Cambrian Intelligence: The Early History of the New AI. Cambridge MA: The MIT Press. ISBN 0-262-52263-2

Buonomano, D. V. & Merzenich, M. M. (1998) Cortical plasticity: from synapses to maps. Annual Review of Neuroscience 21(1): 149-186.

Burgess, P. W. (1996) Confabulation and the control of recollection. Memory 4(4): 359-412.

Campbell, D. T. (1990) Levels of organization, downward causation, and the selection-theory approach to evolutionary epistemology.

Cannon, W. B. (1932) The wisdom of the body.

Carter, J. R. (2000) Facial expression analysis in schizophrenia. Unpublished doctoral dissertation. University of Western Ontario.

Chalmers, D. (1996) The Conscious Mind: In Search of a Fundamental Theory, New York and Oxford: Oxford University Press. Hardcover: ISBN 0-19-511789-1, paperback: ISBN 0-19-510553-2

Clark, A. (2013) Whatever next? Predictive brains, situated agents, and the future of cognitive science. Behavioral and Brain Sciences 36(03): 181-204.

Clark, H. H. (1969) Linguistic processes in deductive reasoning. Psychological Review 76(4): 387.

Conant, R. C. & W. Ashby, R. (1970) Every good regulator of a system must be a model of that system. International Journal of Systems Science 1(2):89-97.

Cowan, N. (2001) The magical number 4 in short-term memory: A reconsideration of mental storage capacity. Behavioral and Brain Sciences 24(1): 87-114.

Craik, F. I. & Lockhart, R. S. (1972) Levels of processing: A framework for memory search. Journal of Verbal Learning and Verbal Behavior 11(6): 671-684.

Crick, F. H. (1958) On protein synthesis. In Symposia of the Society for Experimental Biology (Vol. 12, p. 138).

Crick, F. (1970) Central dogma of molecular biology. Nature, 227(5258): 561-563.

Dalton, P. (2000) Psychophysical and behavioral characteristics of olfactory adaptation. Chemical Senses, 25(4): 487-492.

Damasio, A. (1999) The Feeling of What Happens: Body and Emotion in the Making of Consciousness. New York: Houghton Mifflin Harcourt. ISBN 0-15-601075-5

Darwin, C. (1859/2009) The origin of species by means of natural selection: or, the preservation of favored races in the struggle for life. W. F. Bynum (Ed.). AL Burt.

Dawkins, R. (1999) The extended phenotype: The long reach of the gene. Oxford University Press.

Dawkins, R. (2004) Extended phenotype–but not too extended. A reply to Laland, Turner and Jablonka. Biology and Philosophy 19(3): 377-396.

Day, T. A. (2005) Defining stress as a prelude to mapping its neurocircuitry: no help from allostasis. Progress in Neuro-Psychopharmacology and Biological Psychiatry 29(8): 1195-1200.

Dehaene, S. & Changeux, J. P. (1991) The Wisconsin Card Sorting Test: Theoretical analysis and modeling in a neuronal network. Cerebral Cortex 1(1): 62-79.

Descartes, R. (1983) Principia philosophiae (Principles of Philosophy). Translation with explanatory notes by Valentine Rodger and Reese P. Miller (Reprint ed.). [1644, with additional material from the French translation of 1647]. Dordrecht: Reidel. ISBN 90-277-1451-7.

Di Paolo, E. & De Jaegher, H. (2012) The Interactive Brain Hypothesis, Frontiers in Human Neuroscience 6: 163. http://dx.doi.org/10.3389/fnhum.2012.00163

Draganski, B., Gaser, C., Busch, V., Schuierer, G., Bogdahn, U. & May, A. (2004) Neuroplasticity: changes in grey matter induced by training. Nature 427(6972): 311-312.

Dubner, R. & Ruda, M. A. (1992) Activity-dependent neuronal plasticity following tissue injury and inflammation. Trends in Neurosciences 15(3): 96-103.

Eglen, S. J., & Gjorgjieva, J. (2009) Self- organization in the developing nervous system: Theoretical models. HFSP journal 3(3): 176-185.

Eriksson, D., Valentiniene, S., & Papaioannou, S. (2010) Relating information, encoding and adaptation: decoding the population firing rate in visual areas 17/18 in response to a stimulus transition. PloS one, 5(4), e10327.

Eriksson, D., Wunderle, T., & Schmidt, K. (2012) Visual cortex combines a stimulus and an error-like signal with a proportion that is dependent on time, space, and stimulus contrast. Frontiers in systems neuroscience, 6.

Furmanski, C. S. & Engel, S. A. (2000) Perceptual learning in object recognition: Object specificity and size invariance. Vision Research 40(5): 473-484.

Freud, S. (1915/2005) The unconscious. (G Frankland, Trans.) Penguin, London.

Fris, R. J. (2004) Preoperative low energy diet diminishes liver size. Obesity Surgery 14(9): 1165-1170.).

Friston, K. (2010) The free-energy principle: a unified brain theory? Nature Review Neuroscience 11 (2): 127–38.

Friston, K., Adams, R. A., Perrinet, L. & Breakspear, M. (2012) Perceptions as hypotheses: saccades as experiments. Frontiers in Psychology 3.

Gallese, V. & Lakoff, G. (2005. The brain's concepts: The role of the sensory-motor system in conceptual knowledge. Cognitive Neuropsychology 22(3-4): 455-479.

Ganguly, K. & Poo, M. M. (2013) Activity-Dependent Neural Plasticity from Bench to Bedside. Neuron 80(3): 729-741.

Gibson, J. J. (1979) The Ecological Approach to Visual Perception, ISBN 0-89859-959-8.

Gibson, J. J. (1977) The Theory of Affordances. In Perceiving, Acting, and Knowing, edited by Robert Shaw and John Bransford, ISBN 0-470-99014-7.

Giovannoni, S. J., H. James Tripp et al. (2005) Genome Streamlining in a Cosmopolitan Oceanic Bacterium. Science 309 (5738): 1242–1245. doi:10.1126/science.1114057. PMID 16109880.

Glanville, R. (2002) Second order cybernetics. Encyclopaedia of Life Support Systems.

Gollnick, P., Babitzke, P., Antson, A. & Yanofsky, C. (2005) Complexity in regulation of tryptophan biosynthesis in Bacillus subtilis. Annual Review Genetics 39: 47-68.

Grossberg, S. (1987) Competitive learning: From interactive activation to adaptive resonance. Cognitive Science 11(1): 23-63.

Gurfinkel, V. S., Lipshits, M. I. & Popov, K. E. (1974) Is the stretch reflex a basic mechanism in the system of regulation of human vertical posture?. Biofizika 19(4): 744.

Hansel, C., Linden, D. J. & D'Angelo, E. (2001) Beyond parallel fiber LTD: the diversity of synaptic and non-synaptic plasticity in the cerebellum. Nature Neuroscience 4(5): 467-475.

Hasher, L. & Zacks, R. T. (1979) Automatic and effortful processes in memory. Journal of Experimental Psychology: General 108: 356-388.

Hebb, D. O. (2002) The organization of behavior: A neuropsychological theory. Psychology Press.

Heylighen, F. & Joslyn, C. (2001) Cybernetics and second order cybernetics. Encyclopedia of physical science & technology 4: 155-170.

Holt, E. B. (1914) The concept of consciousness. New York: Macmillan.

Holtmaat, A., & Svoboda, K. (2009) Experience-dependent structural synaptic plasticity in the mammalian brain. Nature Reviews Neuroscience, 10(9): 647-658.

Hubel, D. H. & Wiesel, T. N. (1962) Receptive fields, binocular interaction and functional architecture in the cat's visual cortex. The Journal of Physiology 160(1): 106.

Hurley, J. B. (2002) Shedding light on adaptation. The Journal of general physiology, 119(2): 125-128.

James, W. (1890) The principles of psychology. Henry Holt and Co., New York.

Jerger, J. F. (1957) Auditory adaptation. The Journal of the Acoustical Society of America, 29(3): 357-363.

Johanssen, W. (1911) The genotype conception of heredity. The American Naturalist 45(531): 129-59. http://www.jstor.org/stable/2455747

Jones, K. N., Touchstone, R. M., & Gettys, C. F. (1974) A tactile illusion: The rotating hourglass. Perception & Psychophysics, 15(2): 335-338.

Julesz, B. (1984) A brief outline of the texton theory of human vision. Trends in Neurosciences 7(2): 41-45.

Jung-Beeman, M., Bowden, E. M., Haberman, J., Frymiare, J. L., Arambel-Liu, S., Greenblatt, R., ... & Kounios, J. (2004) Neural activity when people solve verbal problems with insight. PLoS Biology 2(4): e97.

Kahneman, D. (2003) A perspective on judgement and choice. American Psychologist 58: 697-720.

Kahneman, D. (2011) Thinking, Fast and Slow. Macmillan. ISBN 978-1-4299-6935-2.

Kant, I., Guyer, P. & Wood, A. W. (Eds.). (1998) Critique of pure reason. Cambridge University Press.

Kaplan, G. B., Şengör, N. S., Gürvit, H., Genç, İ. & Güzeliş, C. (2006) A composite neural network model for perseveration and distractibility in the Wisconsin card sorting test. Neural Networks 19(4): 375-387.

Karatsoreos, I. N. & McEwen, B. S. (2011) Psychobiological allostasis: resistance, resilience and vulnerability. Trends in Cognitive Sciences 15(12): 576-584.

Kirschner, M. & Gerhart, J. (1998) Evolvability. Proceedings of the National Academy of Sciences of the United States of America 95 (15): 8420–8427.

Kleinschmidt, A., Büchel, C., Zeki, S. & Frackowiak, R. S. J. (1998) Human brain activity during spontaneously reversing perception of ambiguous figures. Proceedings of the Royal Society of London. Series B: Biological Sciences 265(1413): 2427-2433.

Köhler, W. (1929) Gestalt psychology.

Kosslyn, S. M., Digirolamo, G. J., Thompson, W. L. & Alpert, N. M. (1998) Mental rotation of objects versus hands: neural mechanisms revealed by positron emission tomography. Psychophysiology 35(2): 151-161.

Kurzweil, R. (2005) The singularity is near: When humans transcend biology. Penguin. com.

Lakoff, G. & Johnson, M. (1980) Metaphors We Live By. University of Chicago Press. ISBN 0-226-46801-1

Levine, D. S. & Prueitt, P. S. (1989) Modeling some effects of frontal lobe damage—novelty and perseveration. Neural Networks 2(2): 103-116.

Liddell, E. G. T. & Sherrington, C. (1924) Reflexes in response to stretch (myotatic reflexes). Proceedings of the Royal Society of London. Series B, Containing Papers of a Biological Character 96(675): 212-242.

Lieberman, M. D. (2007) Social cognitive neuroscience: A review of core processes. Annual Review of Psychology 58:259–89. 10.1146/annurev.psych.58.110405.085654

Maturana, H. R. & Varela, F. J. (1980) Autopoiesis and cognition: The realization of the living. Vol. 42. Springer.

Maturana, H. R. & Varela, F. J. (1992) The tree of knowledge: the biological roots of human understanding (Rev. ed.). Boston: Shambhala.

Mayer, J. S., Bittner, R. A., Nikolić, D., Bledowski, C., Goebel, R. & Linden, D. E. (2007) Common neural substrates for visual working memory and attention. Neuroimage 36(2): 441-453.

McCloskey, M. & Cohen, N. (1989) Catastrophic interference in connectionist networks: The sequential learning problem. In G. H. Bower (ed.) The Psychology of Learning and Motivation 24: 109-164.

McEwen, B.S. & Stellar, E. (1993) Stress and the individual. Mechanisms leading to disease. Archives of internal medicine 153 (18): 2093–101. PMID 8379800.

McGann, M., De Jaegher, H. & Di Paolo, E. A. (2013) Enaction and psychology. Review of General Psychology 17(2). http://dx.doi.org/10.1037/a0032935

Meehl, P. E., & Sellars, W. (1956) The concept of emergence. Minnesota studies in the philosophy of science, 1: 239-252.

Miller, G. A. (1956) The magical number seven, plus or minus two: some limits on our capacity for processing information. Psychological Review 63(2): 81.

Mitchell, T. M. (1980) The need for biases in learning generalizations (pp. 184-191). Department of Computer Science, Laboratory for Computer Science Research, Rutgers Univ.

Mozzachiodi, R. & Byrne, J. H. (2010) More than synaptic plasticity: role of nonsynaptic plasticity in learning and memory. Trends in Neurosciences 33(1): 17-26.

Nikolić, D., & Gronlund, S. D. (2002) A tandem random walk model of the SAT paradigm: Response times and accumulation of evidence. British Journal of Mathematical and Statistical Psychology, 55(2): 263-288.

Nikolić, D. & Singer, W. (2007) Creation of visual long-term memory. Perception & Psychophysics 69(6): 904-912.

Nikolić, D., Häusler, S., Singer, W., & Maass, W. (2009) Distributed fading memory for stimulus properties in the primary visual cortex. PLoS Biology: 7(12), e1000260.

Nikolić, D. (2010) The brain is a context machine. Review of Psychology 17(1): 33-38.

Noble, D. (2008a) The music of life: biology beyond genes. Oxford University Press.

Noble, D. (2008b) Claude Bernard, the first systems biologist, and the future of physiology. Experimental Physiology 93(1): 16-26.

Noë, A. (2012) Varieties of Presence. Cambridge, MA: Harvard University Press. ISBN 9780674062146.

O'Connor, T. (1994) Emergent properties. American Philosophical Quarterly: 91-104.

Olsson, H. & Poom, L. (2005) Visual memory needs categories. Proceedings of the National Academy of Sciences of the United States of America 102(24): 8776-8780.

Paivio, A. (2007). Mind and its evolution: A dual coding theoretical approach. Mahwah, NJ. Lawrence Erlbaum Associates.

Parks, R. W., Levine, D. S., Long, D. L., Crockett, D. J., Dalton, I. E., Weingartner, H., ... & Becker, R. E. (1992) Parallel distributed processing and neuropsychology: A neural network model of Wisconsin Card Sorting and verbal fluency. Neuropsychology Review 3(2): 213-233.

Peirce, C. S. (1903) Harvard lectures on pragmatism, Collected Papers v. 5.

Popper, K.R. (1999) Notes of a realist on the body–mind problem. All Life is Problem Solving (A lecture given in Mannheim, 8 May 1972 ed.). Psychology Press.

Posner, M. I. & Petersen, S. E. (1990) The attention system of the human brain. Annual Review of Neuroscience 13: 25–42.

Powers, W. T. (1973) Behavior: The control of perception. Chicago: Aldine de Gruyter. ISBN 0-202-25113-6.

Ratcliff, R. (1990) Connectionist models of recognition memory: Constraints imposed by learning and forgetting functions. Psychological Review 97: 285-308.

Robinson, W. S. (2005) Zooming in on downward causation. Biology and Philosophy, 20(1):117-136.

Rosenblatt, F. (1958) The perceptron: a probabilistic model for information storage and organization in the brain. Psychological Review 65(6): 386.

Rummelhart, D. E., Hinton, G. E. & Williams, R. J. (1986) Learning representations by back-propagating errors. Nature 323(9): 533-535.

Rust J. (2009) John Searle. Continuum International Publishing Group. pp. 27–28. ISBN 0826497527.

Schacter, D. L., Norman, K. A. & Koutstaal, W. (1998). The cognitive neuroscience of constructive memory. Annual Review of Psychology 49(1): 289-318.

Shacklock, M. (1995) Neurodynamics. Physiotherapy 81(1): 9-16.

Searle, J. (1980) Minds, Brains and Programs. Behavioral and Brain Sciences 3(3): 417–457, doi:10.1017/S0140525X00005756

Searle, J. (2009) Chinese room Argument. Scholarpedia 4(8):3100

Shiffrin, R. M. & Schneider, W. (1977) Controlled and automatic human information processing: II. Perceptual learning, automatic attending, and a general theory. Psychological Review 84: 127–190.

Shipp, S., Adams, R. A. & Friston, K. J. (2013) Reflections on a granular architecture: predictive coding in the motor cortex. Trends in Neurosciences 36(12): 706-716.

Simon, H. A. (1994) Near decomposability and complexity: How a mind resides in a brain. In: The Mind, The Brain, and Complex Adaptive Systems. Eds.: Harold J. Morowitz and Jerome L. Singer. SFI studies in the sciences of complexity, Vol XXII, Addison-Wesley.

Singer, W., & Tretter, F. (1976) Unusually large receptive fields in cats with restricted visual experience. Experimental Brain Research, 26(2): 171-184.

Singer, P. D. W., Freeman, B., & Rauschecker, J. (1981) Restriction of visual experience to a single orientation affects the organization of orientation columns in cat visual cortex. Experimental brain research, 41(3-4): 199-215.

Sperry, R. W. (1969) A modified concept of consciousness. Psychological review 76(6): 532.

Sperry, R. W. (1980) Mind-brain interaction: mentalism, yes; dualism, no. Neuroscience 5(2): 195-206.

Squire, L. R. (1992) Memory and the hippocampus: a synthesis from findings with rats, monkeys, and humans. Psychological Review 99(2): 195.

Sterling, P. (2004) Principles of Allostasis: Optimal design, predictive regulation, pathophysiology, and rational therapeutics. IN: Schulkin, J. (Ed.). Allostasis, homeostasis, and the costs of physiological adaptation. Cambridge University Press: Cambridge, UK.

Sterling, P. & Eyer, J. (1988) Allostasis: A new paradigm to explain arousal pathology. In: S. Fisher and J. Reason (Eds.), Handbook of Life Stress, Cognition and Health. John Wiley & Sons, New York.

Sternberg R.J., Davidson J.E., eds. (1995) The nature of insight. Cambridge (Mass.): MIT Press.

Szentagothai, J. (1984). Downward causation?. *Annual review of neuroscience*,*7*(1), 1-12.

Szentágothai, J., & Érdi, P. (1989). Self-organization in the nervous system.*Journal of Social and Biological Structures*, *12*(4), 367-384.

Thayer, J. F. & Lane, R. D. (2000) A model of neurovisceral integration in emotion regulation and dysregulation. Journal of Affective Disorders 61(3): 201-216.

Thompson, E. (2007). *Mind in life: Biology, phenomenology, and the sciences of mind*. Harvard University Press.

Treisman, A. (1985) Preattentive processing in vision. Computer Vision, Graphics, and Image Processing 31(2): 156-177.

Treisman, A. M., & Gelade, G. (1980) A feature-integration theory of attention. Cognitive Psychology 12(1): 97-136.

Turrigiano, G. (2012) Homeostatic synaptic plasticity: local and global mechanisms for stabilizing neuronal function. Cold Spring Harbor perspectives in biology 4(1).

Turrigiano, G. G. & Nelson, S. B. (2004) Homeostatic plasticity in the developing nervous system. Nature Reviews Neuroscience 5(2): 97-107.

V. Braintenberg (1984) Vehicles: Experiments in synthetic psychology. Cambridge, MA: MIT Press

Varela, F. J., Thompson, E., & Rosch, E. (1991) The embodied mind: cognitive science and human experience. Cambridge, Mass.: MIT Press.

von Foerster, H. (2003) Cybernetics of cybernetics. In Understanding Understanding (pp. 283-286). Springer New York.

Wiener, N. (1961) Cybernetics: On the Control and Communication in the Animal and the Machine. MIT Press. ISBN 0-262-73009-X

Wise, R.A. (1996) Addictive drugs and brain stimulation reward. Annual Review Neuroscience 19: 319–40.doi:10.1146/annurev.ne.19.030196.001535. PMID 8833446

Xu, T., Yu, X., Perlik, A. J., Tobin, W. F., Zweig, J. A., Tennant, K., ... & Zuo, Y. (2009) Rapid formation and selective stabilization of synapses for enduring motor memories. Nature 462(7275): 915-919.

Ylikoski, J., & Lehtosalo, J. (1985) Neurochemical basis of auditory fatigue: a new hypothesis. Acta oto-laryngologica, 99(3-4): 353-362.

Yoshitake, K., Tsukano, H., Tohmi, M., Komagata, S., Hishida, R., Yagi, T. & Shibuki, K. (2013) Visual Map Shifts based on Whisker-Guided Cues in the Young Mouse Visual Cortex. Cell Reports 5(5): 1365-1374.

| | $T_0$ | $T_1$ | $T_2$ | $T_3$ |
|---|---|---|---|---|
| cybernetic knowledge | ✔ | ✔ | ✔ | ✔ |
| variety | ✔ | ✔ | ✔ | ✔ |
| information | ✔ | ✔ | ✔ | ✔ |
| control | — | ✔ | ✔ | ✔ |
| supervision | — | — | ✔ | ✔ |
| anapoiesis | — | — | — | ✔ |
| homeostasis | — | ✔ | ✔ | ✔ |
| allostasis | — | — | ✔ | ✔ |
| peristasis | — | — | — | ✔ |
| storage of knowledge | ✔ | ✔ | ✔ | ✔ |
| knowledge induction | — | ✔ | ✔ | ✔ |
| knowledge deduction | — | — | ✔ | ✔ |
| knowledge abduction | — | — | — | ✔ |

**Table 1.** Properties of systems that exhibit different number of traverses, from zero traverses ($T_0$) to three traverses ($T_3$). All systems possess cybernetic knowledge and variety. However, with increased number of traverses the adaptive capabilities increase. For example, only a $T_3$-system or higher is able to perform anapoiesis (see text for explanation).

| | Classical ($T_{1/2}$) | Anapoietic ($T_{2/3}$) |
|---|---|---|
| operation | computation | adaptation |
| input and output | symbols | monitor-and-act units |
| relation to surroundings (Gibson; Brooks; Nöe; *supra*) | boxed | permeable |
| logical operation (Peirce, *supra*) | deduction | abduction |
| symbol treatment (Searle, *supra*) | syntax | semantics |
| repeatability | faithful | creative |
| emergence | supervening | non-supervening |
| physiological implementation | electrochemical neural activity | processes of neural adaptation |

**Table 2.** Two different approaches towards answering the question: What is thought? The table makes the contrast between the classical and anapoietic approaches. The classical approach is based on $T_2$-systems, having one traverse for executing mental operations and a separate one for learning. In contrast, anapoietic approach is based on $T_3$-systems, employing in total two traverses for executing mental operations—one traverse for generating behavior and a separate one for mental operations bona fide. The anapoietic approach agrees with several previous theoretical works suggesting external components to cognition, importance of semantics in symbol treatment and operations based on logical abduction. In addition, anapoiesis offers several new insights into the nature of the thought process.

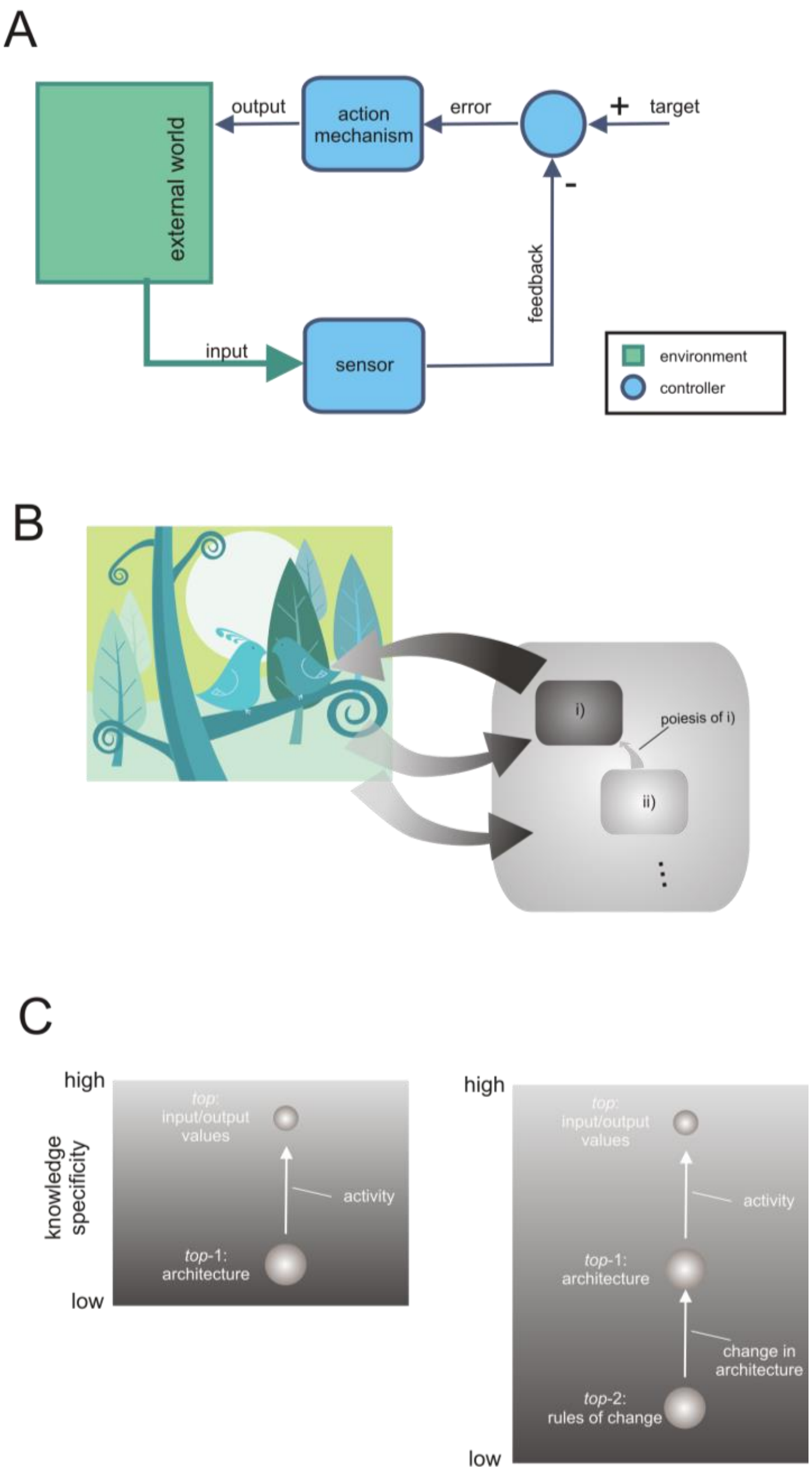

**Figure 1:** Cybernetic systems and the acquisition of cybernetic knowledge through practopoiesis. A) Interaction graph of a classical cybernetic control system implementing monitor-and-act machinery. B) The basic principle of practopoietic acquisition of cybernetic knowledge. If subsystem i) represents a classical cybernetic system like the one in A) and operates at a higher level of organization, the subsystem ii) operates at a lower level of organization to make changes to i) such that i) obtains proper cybernetic knowledge. Actions performed by ii) have poietic effects on i) and for that require feedback from the environment. The three dots indicate that this organizational relationship can be generalized as yet another subsystem may provide cybernetic knowledge for ii). C) Graphs of the relationships in the specificity/generality of cybernetic knowledge or *knowledge graphs*, shown for the components of systems in A) (left) and B) (right). Left: The system exhibits two levels of knowledge, i.e. two levels of organization (spheres): It contains general knowledge about the rules of control in the form of the system architecture, and more specific knowledge about the current states in the form of its input/output values. The arrow indicates the transition i.e., *traverse*, of knowledge from general to specific, which is a function of the operation of the system. Right: The system has one more level of organization and thus, one more traverse. The most general knowledge is that containing the rules for changing system architecture. The levels of organization are indicated by *top*, *top*-1, etc. The relative sizes of spheres indicate the total amount of knowledge stored at each level of organization i.e., its cybernetic variety.

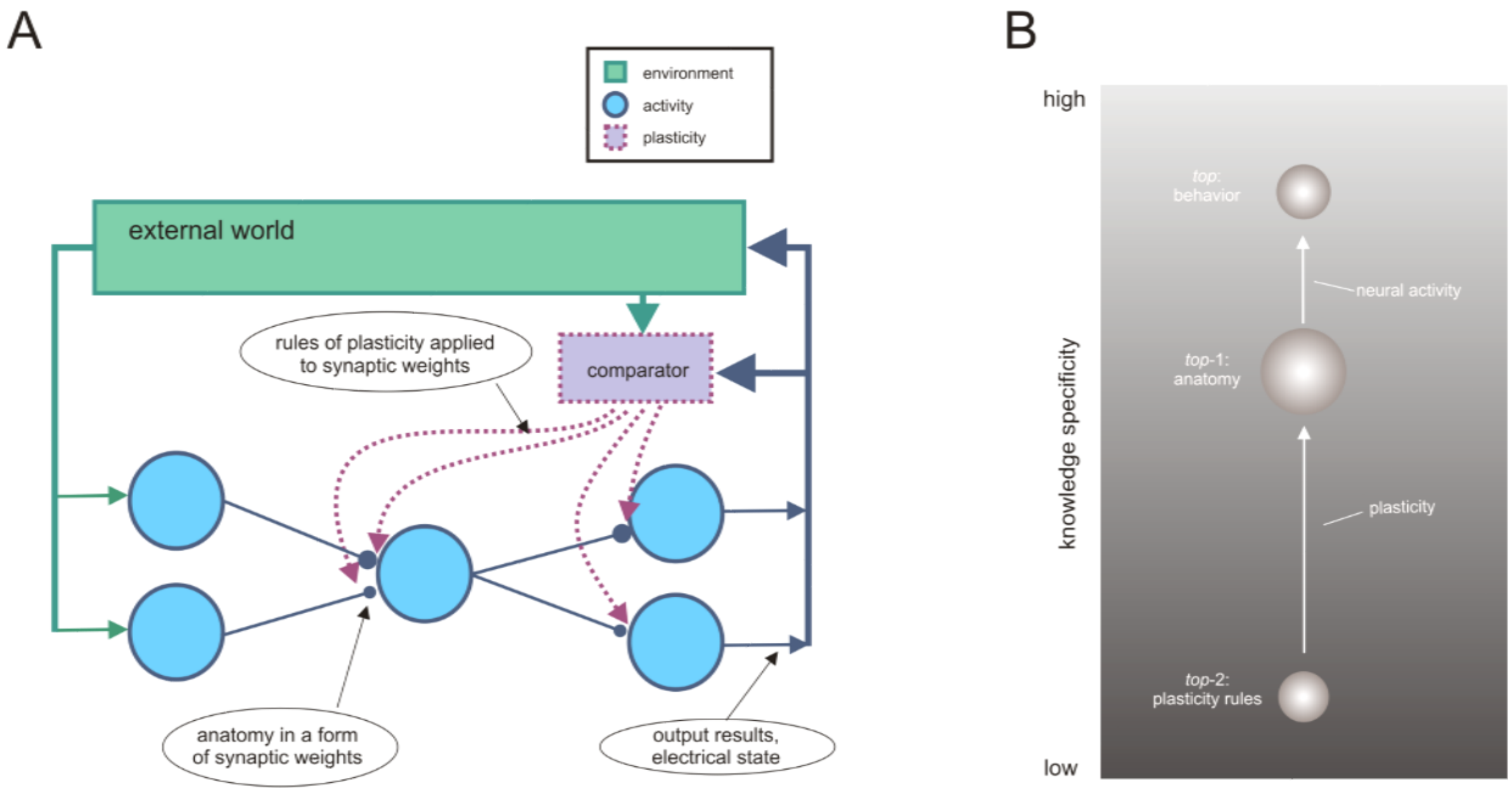

**Figure 2:** Example systems that exhibit one step more adaptability during self-organization than classical cybernetic systems. A) The interaction graph of various components underlying supervised learning in back-propagation and similar algorithms for learning in neural networks. Blue: the top mechanism that implements an input-output function with the environment. Purple: an adaptive mechanism at a lower level of organization that provides cybernetic knowledge for the *top* in a form of synaptic weights. Green: environment from which both mechanisms obtain feedback. B) Adaptive function of plasticity mechanisms in natural neural networks shown as the relationship of the specificity of cybernetic knowledge. There are three levels at which knowledge is stored i.e., three organizational levels, and two types of mechanisms that enable traverse from general to specific knowledge: plasticity rules are needed for creating network anatomy, and network anatomy is needed for creation of behavior.

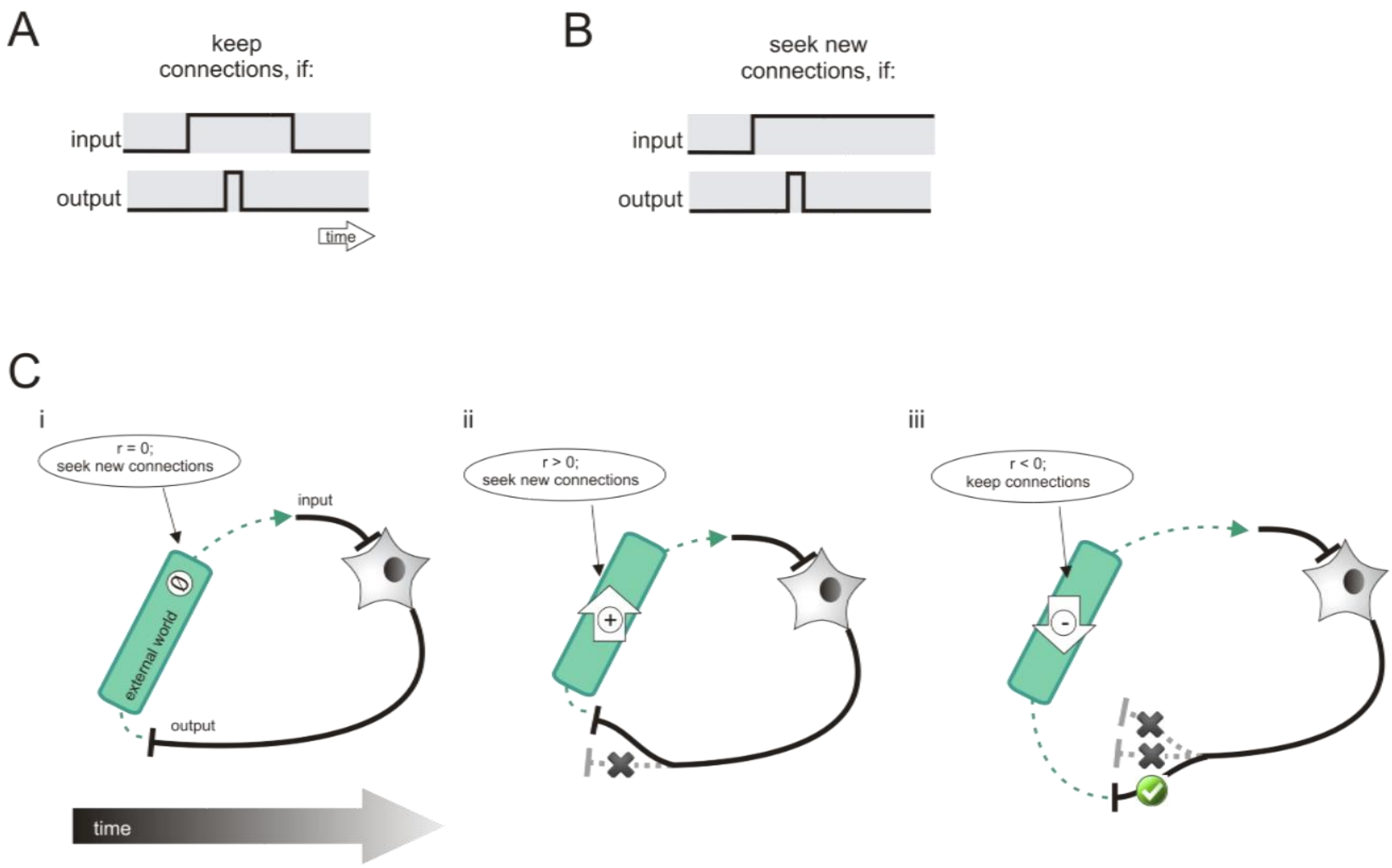

**Figure 3:** Creation (poiesis) of a functional reflex arc in a two-traversal system by applying hypothetical plasticity rules that rely on level-specific environmental feedback. **A)** The plasticity rule to keep or strengthen connections: A contingency is sought in which an input produces output and is quickly followed by a removal of the input. This is taken as an indicator that a neuron's actions remove the transpiring inputs, which is in turn an indicator of a good performance of the reflex. **B)** The plasticity rule to dispose connections and seek new ones: The output of a neuron is not followed by a removal of the input. **C)** A hypothetical sequence of events in the process of poiesis of a monitor-and-act unit at a higher level of organization (reflex) by the actions at a lower level of organization (plasticity rules): i) At first, there is no detectable contingency between input and output. ii) This prompts the neuron to abandon existing synapses and to seek new ones. A new one may produce contingencies but not necessarily a desirable one (e.g., forming a positive rather than a negative feedback loop). iii) A further search for a synapse finally results in a desirable negative feedback loop, and is kept and maintained.

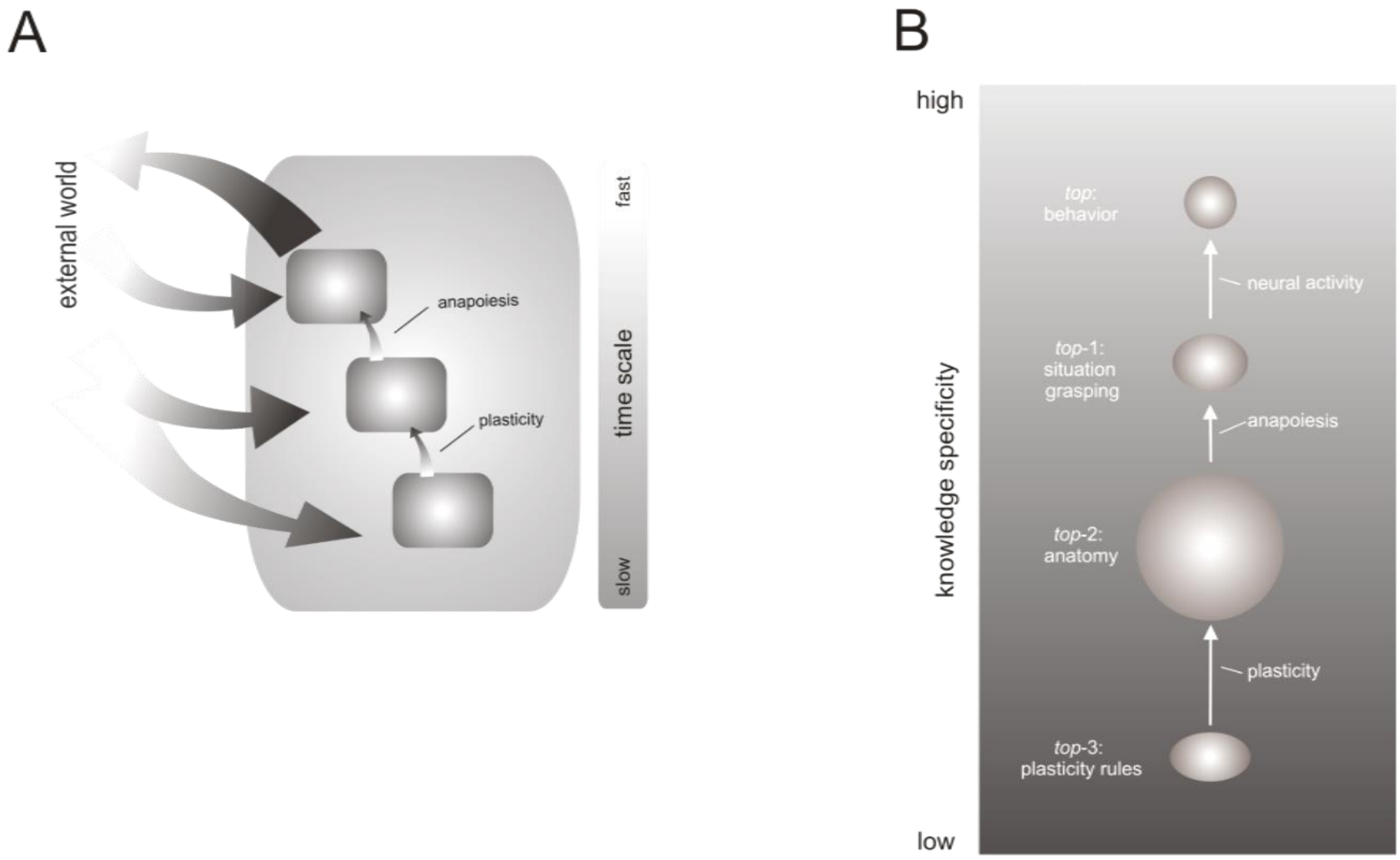

**Figure 4:** A tri-traversal system obtained by inserting a traverse in-between plasticity and neural activity. Such a system implements *anapoiesis* as a middle traverse that lies between plastic changes creating anatomy and neural activity creating behavior. Anapoiesis extracts knowledge about the current situation and may be the missing component needed to account for human cognition. A) An interaction graph indicating that anapoiesis needs its own feedback from environment in order to grasp the current situation. The time scales indicate that situation grasping is mostly a quicker process than extraction of network anatomy, but also a slower process than the sensory-motor loops of neural activity needed to generate behavior. B) The four levels of organization and the corresponding three traverses shows in a knowledge graph. Knowledge extracted by anapoiesis is more specific than the knowledge stored in the anatomy of the system, but is more general than that extracted by neural activity.

[1] According to Bechtel and Richardson 1992: "A system will be nearly decomposable to the extent that the causal interactions within subsystems are more important in determining component properties than are the causal interactions between subsystems"

[2] Monitor-and-act components offer a much more complete descriptor of an adaptive system than the concept of represented information. The latter requires additional mechanisms to be defined for encoding and decoding information. In other words, besides the memory, a mechanism for computation is required—something that has the knowledge of the code. In contrast, a monitor-and-act unit houses both "memory storage" and "processor" under the same roof. It is a complete action system that autonomously performs the entire cycle of detecting information, acting on it and observing the effects of the action. In other words, the symbols providing information require an external grounding process. In contrast, a monitor-and-act unit possesses already by itself everything necessary to operate.

[3] The practopoietic process is very much different from building a system from a pre-defined plan. When building predefined systems no feedback is needed. In contrast, during the knowledge acquisition of poiesis low-level components receive environmental feedback that necessarily guides the creation of new structures.

[4] High and low levels of organization in practopoiesis may seem counterintuitive and they may stand in opposition to traditional definitions of hierarchies in the brain whereby the highest element is the one that has the most decision power. These are hierarchies of power control. In contrast, practopoiesis is a hierarchy of organization levels. In practopoiesis, the more highly organized component is the one that cannot operate without the guidance of the lower one. For example, one may see genes as having very high decision power, higher than e.g., behavior: Genes can determine behavior and behavior cannot directly determine genes. Nevertheless, genes operate at a lower level of system organization than does behavior. Similarly, in a management structure of a social organization, the highest decision power (e.g., a CEO) does not coincide with the highest level of system organization. The result of the collective action of all the employees may constitute a much higher form of organization than the guiding actions of the CEO. The main concern of practopoietic theory is the level of organization, rather than the decision power.

[5] Note that what is considered as the *top* organization level of a system is relative and depends on what has been chosen to be considered as a system. For example, if an organism is considered as an adaptive system, then its *top* level corresponds to its behavior and the environment with which it interacts corresponds to the world surrounding the organism. But if a cell within that organism is the object of the analysis or an organ of an organism, then the *top* levels corresponds to the functions of those cells/organs and the environments with which they interact include other cells and other organs within the organism. Similarly, a system that transcends a single organism can be the object of the analysis, such as for example, a human+machine system, or a social organization consisting of multiple human members. In those cases the top levels of organization may be different from those of individual organisms and so may be different the environments with which those systems interact.

[6] Hence *top* level may exert poietic effects possibly only on its environment.

[7] Traverse takes place already at the lowest levels of system organization. For example, the process of evolution by random change and selection is a traverse too: A general knowledge of evolvability (Kirschner & Gerhart 1998) is used to extract a more specific knowledge about the actual evolved properties of the system (e.g., a genome). Whether an individual successfully procreates or not is a form of feedback obtained from the environment, and requires involvement of the top level of organization i.e., behavior. That way, cybernetic knowledge can be acquired through evolution and stored in the genome and organelles.

[8] For example, an artificial neural network may implement a large number of different input-output mapping functions by applying learning rules. The variety needed to implement learning rules is much smaller in comparision to the total number of input-output functions that such a system can in principle learn. Another, even more extreme example is evolution by natural selection whereby application of a single set of rules—random change combined with natural selection—produced over long period of time an entire kingdom of living forms on the planet earth.

[9] This is similar to the difference between a lookup table (fast access but large storage resources) and computation on the spot (slower access but a leaner and more flexible system). Thus, instead of a series of if-then statements, as in an expert system, an adaptive system relies on general rules applied in each situation de novo in order to help infer the next action. A key difference to computer algorithms is that most of the variety in computer software is generated by traverses executed by the brains of human operators. Intelligent adaptive systems do not have this external help but have to adjust on their own.

[10] Although, it would be possible, in theory, to equip a single-traverse system with all the necessary knowledge for all the possible events, this works well only for simple artificial environments. Under real-life conditions, as are the survival conditions for a mammal on the planet earth, the combinatorial explosion of the number of possible situations that the system may encounter is too large. Consequently, the system has to rely on abstract rules and extraction of knowledge at multiple levels of organization and thus, on the use of multiple traverses.

[11] The number of levels of organization in a system that are prominent and that play an important role in the system will necessarily be small. There are good reasons for this: Each level requires excessive resources and extensive knowledge acquisition. To be influential, an organization level must possess large cybernetic variety. There is always a possibility that certain parts of the system organize themselves into an even large number of levels of organization and thus, form deeper adaptive structures. However, these parts of the system may rely on small variety and hence, play a relatively small role in the overall adaptability of the system. In other words, to near-decompose the system effectively, we are interested in a small number of organization levels that account for most of the system's cybernetic variety.

[12] Note that the processing stages of neural networks (such as layers of a perceptron (Rosenblatt 1958), or stages of the processing streams in the cortex (Hubel & Wiesel 1962) form a hierarchy that is different from that of practopoietic systems. The membranes of two neurons may operate poietically at the same level, although one neuron may be located at a higher brain area than the other. For example, a cortical neuron is phylogenetically and ontogenetically higher than a spinal neuron and yet, the inhibition/excitation mechanisms of the two operate equi-level, while their respective plasticity mechanisms lie at a practopoietically lower level, and they both also operate equi-level. Similarly, the "classical" hierarchy of processing stages in vision: retina>LGN>V1>V2>…>IT is not a poietic hierarchy.

[13] One consequence of multiple levels of organization is that often an adaptive system can neither be built quickly nor can it make large adjustments quickly. Instead, the system must proceed in steps of small changes, each being subjected to verification and further adjustments. Much like the evolution of complex species can occur only slowly (Darwin 1859), the growth and learning of a highly organized organism (or any other adaptive intelligent system) must progress in small steps. New structures are built gradually on top of the existing ones. Extensive changes require time.

In case that the situation does not allow time for changes, or the changes cannot be achieved at all, the system can be said to experience stress. The system makes changes that are not fully optimized and not enough time is given to reach the best possible balance of all cybernetic knowledge. Some functionality (i.e., health) is necessarily sacrificed. In worst case, the system may not be successful; an organism dies prematurely, or a species gets extinct.

[14] An additional organization level may be provided below plasticity by the so-called phenomenon of metaplasticity, or plasticity-of-plasticity (e.g., Abraham & Bear 1996). However, currently it is not known whether these mechanisms involve level-specific environmental feedback at all the levels of organization (i.e., at all levels of plasticity), which would be required in order to qualify as a multi-level practopoietic system.

[15] Note that the total count of cells alone is not sufficient to produce all the necessary variety. The cells need to be equipped also with the correct cybernetic knowledge. The content of that knowledge is crucial in determining the total intelligence of the system. This is because systems may differ in how good models of the surrounding world they. This is why a human, although equipped with a smaller brain than e.g. a whale, can exhibit in many aspects more intelligent behavior than a whale.

[16] Mechanisms responsible for a positive feedback loop can be equally so considered as $T_1$. For example, mechanisms that evoke emotions may evoke behavior that further intensify the same emotions, acting thus in a positive feedback loop (Thayer & Lane 2000). In those cases monitor-and-act components make a certain response progressively step-up and these increases may be of equal importance for the survival of an organism as negative feedback-based homeostatic regulations.

[17] Devices that we build are set to interact with the environment by producing a single traverse—often, a part of the interaction is a human operator/user. For example, a TV-set is made to interact with its human environment as to get inputs (through button presses) and to deliver outputs (sound, picture). With very few exceptions, our technology improves through an increase in cybernetic variety, not through an increase in the degree of adaptability. That is, the machinery is not being added new traverses. Rather, the number of different responses across different situations is increased. New circuitry is added by human engineers, not by the system that would find ways to improve by itself. Hence, von Neumann computer architecture is used almost exclusively as a high-variety but not a high-adaptability system—keeping its operations mostly at the $T_1$-level.

[18] While in the pre-industrial era $T_0$-artifacts dominated the human civilization in forms of various energy-passive objects such as tools, books, houses, and cold weapons, the industrial and information era brought extensive use of energy-consuming devices and thus, proliferation of $T_1$-systems. Any other adaptive needs that exceed $T_1$ rely mostly on human operators, whose minds operate with more traverses.

Similarly, our formal mathematical tools for scientific and engineering descriptions are mostly suitable for describing operations of $T_1$-systems. We use an equation to make inferences and decisions, the application of which is often a $T_1$-system. The system describes a traverse from a general rule specified by the equation to the specifics of input and output values. By that token a hand-held calculator is a $T_1$-system, and equally so is a complex calculation implemented in a spreadsheet software. In general, a formal logical system with premises and conclusions is a $T_1$-system, whereby a single traverse suffices to derive a conclusion from the premises. Whereas the discipline of mathematics requires much more than $T_1$ for creative formulation of problems and insights on possible solutions, in the end, solutions and proofs are reduced down to a set of $T_1$ operations. Whenever the human mind operates logically, its high-level adaptive capabilities are reduced to much less adaptive (but usually more reliable) $T_1$-operations.

[19] This number could easily exceed the number of atoms in the universe. Hence, it is not possible to device a physical storage or the needed information, not to mention the impossibility of conceiving a mechanisms by which this knowledge would be acquired. For example, there would be not enough time in the age of the universe to acquire such knowledge by a process of evolution by natural selection.

[20] The so-called Hebbian learning mechanisms (Hebb 2002), are not likely to contribute alone considerably the adaptability levels of the system i.e., to the anapoiesis. There are two possible limitations. First, the type of information that they consider is not highly sensitive to environmental influences. For example, spike-timing dependent plasticity (Bi & Poo 1998), which is a form of Hebbian learning, detects the timing relationships between pre- and post spike-synaptic spike. It is unlikely that this timing provides information about the properties of the environment (Turrigiano & Nelson 2004). Hence, Hebbian learning is not designed for closure of practopoietic cycle of cautions with the external events. The second limitation of Hebbian mechanisms is that it is not clear whether they are capable of altering their properties if the circumstances in the environment require so. For example, if the environment's properties change, thus invalidating the application of Hebbian rules (e.g., maybe now the organism would do better by applying anti-Hebbian learning instead), there is no way for the system to adjust to that change in the environment.

The primary benefit that Hebbian learning mechanisms provide is variety; A system with a Hebbian learning mechanism can produce a larger variety of behaviors than the same system lacking such a learning mechanism. A network that forms a $T_1$-system with an addition of a non-adaptive learning mechanism remains a $T_1$-system. Learning rules other than Hebbian may be more suited for achieving adaptability of the system. For example, a hypothetical mechanism based on timing relations between input and output spikes described in Figure 3 is in a better position to use feedback from environmental than is Hebbian learning.

[21] For example, a learning rule (*top*-2) such as the one in Figure 3 may be implemented to detect that operations are suboptimal, which then results in changes made at the level of anatomy (*top*-1). These changes in turn affect how behavior is produced (*top*) in order to satisfy the needs of the organism. Thus, an allostatic change at one level (e.g., network anatomy) keeps another variable constant (e.g., supply of nutrition to the organism).

[22] In human-made machinery, implementations of $T_2$-systems exist but are rare and almost exclusively limited to control based on a single variable. Thus, at the level of supervision, these $T_2$-systems exhibit a minimum of cybernetic variety. Examples include various servomechanisms—e.g., temperature (car engine, computer processor), speed (cruise control) or angular position (robotic arm). These

mechanisms usually ensure proper functioning of another process that itself closes a loop with the environment and can be described as $T_1$ (the operations of the car, computer, robot), making it in total a $T_2$-system. To the best of the author's knowledge, no variety-rich engineered $T_2$-system exists. That is, no artificial system with two traverses contains extensive cybernetic knowledge at the lower traverse. Another example of high-hierarchy but relatively low-variety interaction systems may be management hierarchies in social organizations, in which usually only a few instructions are given by the person in charge as a supervisor, and by the supervisor of a supervisor, and so on (note the reversal of practopoietic and social hierarchies, note 22).

[23] One commonly encountered problem in pattern recognition is the so-called *stability-plasticity dilemma* (Grossberg 1987), also known as *catastrophic interference* (McCloskey & Cohen 1989; Ratcliff 1990). As $T_1$/$T_2$-networks learn new datasets, they forget the old ones. The level *top*-1 forgets what it has known earlier. For $T_2$-systems, to acquire two different types of responses for two different datasets, their samples need to be intermixed, avoiding any temporal grouping—a requirement in discord with the real life.

[24] To reconstruct that knowledge at the *top*-1 level of organization and to use it for controlling behavior, the system still needs to interact with the environment, but this time in a more efficient way. The system may initiate a poietic process from a simple hint from the environment and then seek level-specific environmental feedback about the efficiency of its reconstructive operations again from a relatively modest and quickly accessible environmental sources.

[25] A $T_2$-system i.e., a system without anapoiesis, cannot adjust to new properties of the environment—an equivalent of a new situation—without forgetting how it had adjusted to the previous properties of the environment. To be able to enter a new dataset abruptly and to adjust *top*-1 accordingly at a momentary notice, an elaborate *top*-2 must be developed. Hence, a $T_3$-system exhibiting anapoiesis may be required.

[26] The development of an organism is an anapoietic process. The process of evolution can be considered as a traverse–i.e., the lowest traverse of a species-individual system. If the individual has at least an adaptability level of $T_2$, the species as a whole, including its evolution, exhibits the adaptive level of $T_3$ (and if the organism is $T_3$, as argued later in the text, the species as a whole would be $T_4$). This means then that the traverse from genes to the anatomy of the system exhibits anapoiesis. For example, the growth of an oak tree is an anapoietic process from general knowledge on how to build an oak tree stored in genes to the actual instantiation of the oak. This re-construction process involves interaction with the environment and thus, the actual outcome is somewhat different in each instance, depending on the properties of the environment. For example, to maximize the amount of exposure to sunlight, shapes of branches and orientations of leafs may vary depending on the given situation. The same holds for the growth of other organisms. The anapoiesis of the organism depends on the environment. For example, the animal's fur may grow thicker in colder environment, or the liver can grow larger for a certain diet (Fris 2004).

Anapoiesis is responsible for different phenotypes given the same genotype. Phenotype does not come from a pre-determined plan in genes. Genes create structure through feedback processes and regulation. Thus, the amounts of various created structures depend largely on the properties of the environment. In identical environments, identical phenotype would be obtained for the same genotype, but when properties of the environment vary, as in the real life, also phenotypes necessarily vary. Perhaps the most sensitive aspect of our anatomy to environmental factors, the one that is made to be susceptible as much as possible, are the anatomical structures that store our long-term memories. Everything that we learn is in effect a phenotype of the learning rules stored in the genotype.

Hence, also anapoiesis is the schooling-based transfer of knowledge from one generation to the next. Parents can educate their offspring by conveying knowledge not transferable through genes. This may include learning how to open a nut, hunt, use language, or appreciate music. This knowledge, provided during upbringing, is necessarily stored at higher levels of system organization than genes. Our culture is essentially anapoietic—it has to be re-created in every new individual. It is a form of phenotype (or extended phenotype; Dawkins 1999; Dawkins 2004). Anapoiesis makes it possible to skip the tedious process of rediscovering knowledge from scratch. Instead, a poiesis of specific knowledge is propelled by combining the inborn general cybernetic knowledge with a stimulating environment provided by the environment.

[27] One question is how far can the hierarchy of adaptability go and thus, can $T_5$ or larger systems exist? At present, no evidence for use of such larger systems to any significant extent seems to exist. For a $T_5$-system we should find an additional organizational levels in either of the two directions, either one that is more fundamental than the mechanisms of Darwinian evolution, or one that lies organizationally above our behavioral actions.

A putative mechanism that controls evolution would be much slower than the process of evolution itself and would exert control on how evolution is executed. For as far as the author's knowledge of biology reaches, there seems no evidence to exist that the process of natural evolution is controlled by another mechanism that possesses even more general cybernetic knowledge. However, this does not mean that one should not keep one eye open towards this possibility.

A $T_5$-system created by adding another level of organization on top of the existing biological hierarchy would have to involve some form of development of technology. In that case, our behavioral acts would create machines and these machines would then act on the environment. However, most of our technology cannot be considered as representing another level of practopoietic organization because our interactions with the environment do not exclusively depend on these machines. We usually use machines in parallel with direct behavioral actions. Therefore, interactions machines-humans are mostly equi-level. Nevertheless, it is not difficult to create a scenario, e.g. an underwater robot, whereby in principle, and for a limited environment and a limited period of time, the interactions completely depend on the monitor-and-act units in form of machines created by humans. Thus, for this limited case it could be said that the combined system *human* + *robot* has practopoietic hierarchy larger than $T_3$.

Importantly, however, if human + machine make a $T_4$-system, it does not mean that *human species* + *machine* automatically make a $T_5$-system. For a practopoietic system to exhibit a certain degree of adaptability, all components of adaptive processes must actually take place. This means that the lowest component of the system (i.e., evolution by natural selection) has to affect the top level (the design of machine). It does not seem that our technology affects the environment such that downward pressure for adjustment is exerted all the way down to evolution of the species. In contrast, our medical technology seems to significantly reduced the pressure exerted on natural selection. Hence, technology seems to have a tendency to serve as an alternative to adaptation by natural selection. That is, by increasing our knowledge at the level of ideatheca we become more competent as a species, resulting in less downward pressure on Darwinian selection for the species.

However, one can imagine a hypothetical scenario in which technology and evolution would work both together to enable survival of the species. In an Armageddon scenario in which human race has to evacuate Earth and has to rely heavily on technology to survive on other planets for which we did not evolve, the species may find itself under strong pressure for evolving due to the limitations in the degree to which technology can simulate the original conditions of the planet earth. In that case a full $T_5$ adaptive system or even higher would commence.

[28] Miller (1956) was the first to point to the reconstruction in working memory (or short-term memory) from long-term memory. He noticed that the memory capacity for a string of letter was higher if it contained familiar combinations of letter (e.g., IBMFBIKGB) than if it was completely random. He referred to this process of organizing the stimulus as "chunking" and noted that it required existing knowledge already stored in long-term memory. According to practopoietic theory, this is the process of anapoiesis from long-term memory to working memory. Later, similar properties have been shown for working memory storage in vision (Alvarez & Cavanagh 2004; Nikolić & Singer, 2007), and that visual chunking cannot be made by combining any raw individual visual features but categories of objects must exist in long-term memory (Olsson & Poom 2005). Thus, the limitations in the capacity of working memory seem to be limited by what can be reconstructed from long-term memory.

[29] The distinction between automatic and controlled processes is the most pervasive dichotomy in psychological science. This distinction has been rediscovered multiple times and has been characterized under different names, such as automatic vs. controlled processes (Shiffrin & Schneider 1977), System 1 vs. System 2 (Stanowich & West 2000), intuition vs. reasoning (Kahneman 2003, 2011), verbal vs. non-verbal (Paivio 2007), bottom-up vs. top-down (Posner & Petersen 1990), pre-attentive vs. attentive (Julesz 1984; Treisman 1980, 1985), unconscious vs. conscious (Freud 1915/2005), dual processes (James 1890), effortless vs. effortful (Hasher & Zacks 1979), and reflexive vs. reflective (Lieberman 2007). The common property of automatic, intuitive, pre-attentive processes is that they are fast, require little attention, exhibit high processing capacity, and are resilient to disturbances. These processes are also associated with little experience of effort. Their main shortcoming is a relative lack of flexibility. When we need a new type of behavior that has not been executed or well-learned in the past, we engage controlled attentive processes and reasoning. These mental activities complete tasks with slower pace, exhibit less processing capacity, are more prone to error in case of distraction, and are associated with conscious experience of effort. For example, driving a car to work may be automatic and effortless, but driving in a new city may require focus and attention.

[30] Attention shares a lot of properties and resources with working memory (e.g., Baddeley 1993; Awh & Jonides 2001; Mayer et al. 2007) so that it is often not clear whether these are two separate phenomena or just different sides of the same phenomenon. For example, the larger the working memory capacity for a certain type of stimuli, equally so much faster is the visual search for those stimuli in an attentional task (Alvarez & Cavanagh 2004). According to the practopoietic theory, the shared mechanism behind these phenomena is the anapoietic reconstruction of knowledge.

[31] Searle illustrates this by putting a human into a hypothetical situation in which the person follows blindly rules for mapping from input set of Chinese characters to output set of Chinese characters. Importantly, the person is not Chinese speaking. Thus, by making these rules elaborate enough and following them accurately, an outside observer may have the appearance that the persons is acting intelligently and is thinking, whereas in fact, the person has no understanding whatsoever of the context of the messages. Searle concludes that this proves that such programmed symbolic rules, although possibly appearing to generate intelligent behavior from outside, cannot be sufficient to produce an AI system capable of human-like understanding and thinking.

Notably, it follows from practopoietic theory that such a rule-based system is not even possible to program in practice for real-life problems but only for simplified toy problems. The insurmountable limitation is in the total amount of cybernetic knowledge that would need to be stored. The total number of possible situations i.e., possible sentences to be answered intelligently, using Chinese characters is too large to be programmed by rules and if stored in a $T_1$-system, the requirements for the amount of needed memory storage may exceed the size of the universe. Thus, human-like level of intelligence can be achieved only if the system stores knowledge in a sufficiently generalized form, can extract and adjust this generalized knowledge on its own, and has the capability of applying it to specific situations. This requires a $T_3$-system.

[32] One possibility that my come to mind is that anapoiesis is implemented by specialized cortical areas. For example, areas located high on the cortex hierarchy, such as the frontal cortex or infero-temporal cortex, may be speculated to supply the anapoietic function by determining the processing of information in classically lower brain areas e.g., sensory and motor cortices. There are several reasons why this hypothesis cannot stand. First, adaptability offered by a tri-traversal system is equally needed for all species that have nervous systems, including those that do not even have a cortex. Therefore, this hypothesis would imply that all the other brains, without these higher areas, are $T_2$-systems—which contradicts basic postulates of practopoietic theory. Second, the general anatomy of the cortex does not satisfy the prerequisite of uni-directional poietic relation. Anatomically, all cortical areas are nearly identical and are reciprocally connected. This arrangement does not allow for transcendence of cybernetic knowledge (higher level being a specific case of more general knowledge at the lower level) and shielding the knowledge at lower levels from the dynamics at higher levels. Thus, while higher brain areas make humans more capable of cognition in comparison to other species those brain areas do not make a difference between $T_2$- and $T_3$-systems. Rather than adding a traverse, the additional brain areas enable unique cognitive abilities of humans by adding cybernetic variety.